\documentstyle[graphicx,aps,multicol]{revtex}

\draft

\begin{document}

\title{Entropy of thermal quasiparticles in nuclei}

\author{M.~Guttormsen\footnote{Electronic address: magne.guttormsen@fys.uio.no}, M.~Hjorth-Jensen, E.~Melby, J.~Rekstad, A.~Schiller and 
S.~Siem}
\address{Department of Physics, University of Oslo, P.O.Box 1048 Blindern, 
N-0316 Oslo, Norway}

\maketitle

\begin{abstract}
Information on level density for nuclei with mass numbers $A\sim$ 20--250 is 
deduced from discrete low-lying levels and neutron resonance data. The odd-mass
nuclei exhibit in general $4-7$ times the level density found for its 
neighboring even-even nuclei at the same excitation energy. This excess 
corresponds to an entropy of $\sim 1.7$~$k_B$ for the odd particle. The 
value is approximately constant for all mid-shell nuclei and for all ground state spins. For these nuclei it 
is argued that the entropy scales with the number of quasiparticles. 
A simple model based on the canonical ensemble theory accounts qualitatively for the observed properties.
\end{abstract}


\begin{multicols}{2}

\section{Introduction}

Nuclear structure at low excitation energy depends on the residual long-range 
two-body interaction. One of the most exciting consequences of this interaction 
is the forming of $J=0$ nucleon pairs, the so-called Cooper pairs, where 
nucleons are moving in time reversed orbitals. It has been a great challenge to
observe the breaking of these Cooper pairs and the general quenching of pair 
correlations as function of temperature $T$ and angular frequency $\omega$. 
Rotational high spin states are essentially described by quasiparticles coupled
to the rotating underlying core. Backbending phenomena and spin alignments have
been described according to their quasiparticle properties.

The concept of thermal quasiparticles is based on the idea that thermal 
properties at low excitation energy are governed by a few quasiparticles 
coupled to the cold core of Cooper pairs. These quasiparticles are thermally 
scattered on available single particle states. Thus, they are not well-defined 
by one single (Nilsson) orbital with given spin and parity, but 
exhibit an average of the spectroscopical properties of the orbitals in the 
vicinity of the Fermi surface. At higher temperatures ($T>0.5$~MeV), the 
pair correlations are quenched and the core of Cooper pairs is no longer well 
defined \cite{Schi1}.

The entropy is a fundamental quantity in the description of many body systems. 
For hot nuclei the entropy has the same importance as the spin alignment has 
for rotational nuclei. It can be evaluated from mechanical variables like 
energy, particle number and volume. Thus, it can be deduced in the 
microcanonical ensemble, which is the appropriate ensemble for an isolated 
system like the nucleus. The microcanonical entropy is defined for all isolated systems, from few-body systems up to infinite systems. It has been shown that phase transitions in small systems can be studied using the microcanonical entropy without invoking the thermodynamical limit \cite{GV99}. 

Although less appropriate, also the canonical ensemble theory is widely applied to nuclei. One advantage of the canonical ensemble is that the thermodynamical observables are smooth functions of temperature due to the Laplace transformation involved in the calculation of the canonical partition function. Therefore, we will model our findings in this work within the canonical ensemble theory.

In a recent paper \cite{GB00}, we extracted the microcanonical single 
quasiparticle entropy in hot $^{161,162}$Dy and $^{171,172}$Yb nuclei to be 
$1.70(15)$ in units of the Boltzmann constant $k_B$. Furthermore, the 
single quasiparticle entropy was found to be approximately constant in the 
1--6~MeV excitation region.

In this paper we will investigate the global systematics of nuclear entropy as 
function of nuclear mass, excitation energy (or temperature), number of 
excited quasiparticles and ground state spin. In Sect.~II we sketch a simple model for thermal 
quasiparticles in nuclei. The method of determining level density by counting 
levels and using neutron resonance spacings is described in Sect.~III. In Sect.~IV we 
display the systematics of entropies and in Sect.~V the properties of thermal 
quasiparticles are discussed. In Sect.~VI we present an application involving thermal quasiparticles and, finally, concluding remarks are given in Sect.~VII.

\section{Simple model for thermal quasiparticles}

In this section we will describe a simple model demonstrating the 
properties of thermal quasiparticles, which in turn will be compared to 
experimental data. The main task is to describe the partition function $Z$ in 
the canonical ensemble for a given temperature $T$ and number of thermal 
quasiparticles $n$. The thermodynamical quantities of interest can be deduced 
from the Helmholtz free energy 
\begin{equation}
F(T)=-T\ln Z(T),
\end{equation}
where we express the temperature $T$ in units of MeV and $k_B$ is set to unity.
From this bridge equation connecting statistical mechanics and thermodynamics, 
we may calculate the entropy $S$, the average excitation energy 
$\langle E\rangle$, the heat capacity $C_V$, and the chemical potential $\mu$ 
at the temperature $T$ by
\begin{eqnarray}
S(T)&=&-\left(\frac{\partial F}{\partial T}\right)_V\\
\langle E(T)\rangle&=&F+TS\\
C_V(T)&=&\left(\frac{\partial\langle E\rangle}{\partial T}\right)_V\\
\mu(T)&=&\frac{\partial F}{\partial n}, \label{eq:muT}
\end{eqnarray}
where $n$ is the number of thermal quasiparticles.

\begin{figure}
\includegraphics[totalheight=17.5cm,angle=0,bb=0 80 350 730]{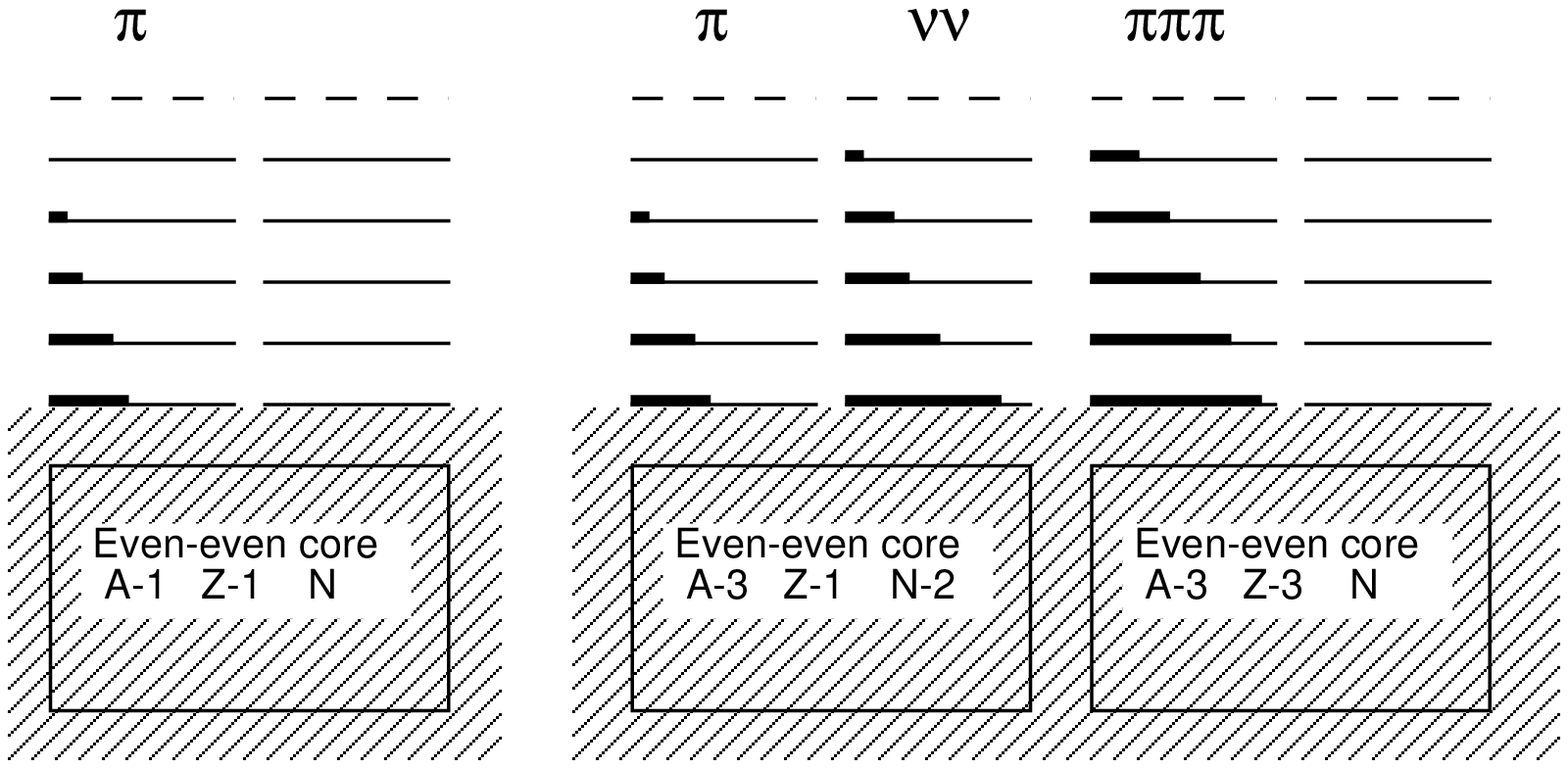}
\caption{Illustration of thermal excitations in an odd proton nucleus, where the horizontal bars indicate how much the orbitals are filled on the average. In the case of no broken pairs (left diagram), the single proton has maximum number of quantuum states (multiplicity) available. The three quasiparticle excitation has to include either a $\pi\nu ^2$ or a $\pi ^3$ configuration (two right diagrams). Here, the Pauli blocking reduces the multiplicity and gives a lower quasiparticle entropy.}
\label{fig:fig1.ps}
\end{figure}

In the evaluation of the partition function $Z$, we assume spin $1/2$ fermions 
scattered into a doubly degenerated single particle level scheme with equal energy spacing $\epsilon$. The level scheme has infinitely many levels, and it is one scheme for protons and one for neutrons, as indicated in Fig.~\ref{fig:fig1.ps}. These conditions simulate the  Nilsson single particle scheme for deformed nuclei.

We will first construct the partition functions $z_n^{\uparrow}$ for $n$ 
identical spin-up fermions placed in one and the same single particle scheme. We assume that 
$n$ quasiparticles are available, with the core of Cooper pairs as a reservoir. The total 
partition functions can then be expressed by these basic partition functions. 
The scattering of one spin-up fermion into the level scheme gives the partition 
function
\begin{equation}
z_1^{\uparrow}=\sum_{i=0}^{\infty}\exp(-i\epsilon/T)
=\frac{1}{1-\exp(-\epsilon/T)}.
\end{equation}
When two identical spin-up fermions are placed into the same level scheme, the 
summing is restricted due to the Pauli principle by
\begin{eqnarray}
z_2^{\uparrow}&=&\sum_{i=0}^{\infty}\left[\exp(-i\epsilon/T)
\sum_{j=i+1}^{\infty}\exp(-j\epsilon/T)\right]\nonumber\\
&=&\frac{1}{1-\exp(-2\epsilon/T)}\frac{\exp(-\epsilon/T)}{1-\exp(-\epsilon/T)}
\nonumber\\
&=&z_1^{\uparrow}\frac{\exp(-\epsilon/T)}{1-\exp(-2\epsilon/T)}.
\end{eqnarray}
Generally, for $n$ identical spin-up fermions we find
\begin{equation}
z_{n}^{\uparrow}=z_{n-1}^{\uparrow}\frac{\exp(-(n-1)\epsilon/T)}
{1-\exp(-n\epsilon/ T)}.
\end{equation}

We now allow spin up {\em and} down for the fermions, and evaluate the 
corresponding partition function $z_n$. For one fermion with spin up {\em or} 
down we simply obtain a degeneration of two, since 
$z_1^{\uparrow}=z_1^{\downarrow}$:
\begin{equation}
z_1=z_1^{\uparrow}+z_1^{\downarrow}=2z_1^{\uparrow}.
\end{equation}
If two fermions occupy the same level scheme, the partition function becomes more 
complicated. With the spins of the two fermions antiparallel ($m=m_1+m_2=0$), 
no Pauli blocking is present and we obtain
\begin{equation}
z_2(m=0)=z_1^{\uparrow}z_1^{\downarrow}=\left(z_1^{\uparrow}\right)^2.
\end{equation}
The contribution with parallel spin invoke the Pauli principle giving
\begin{equation}
z_2(m=\pm 1)=z_2^{\uparrow}+z_2^{\downarrow}=2z_2^{\uparrow}.
\end{equation}
It is now straightforward to evaluate $z_n$ for $n$ fermions, allowing spin up 
and down. For convenience we list the seven lowest partition functions here:
\begin{eqnarray}
z_1&=&2z_1^{\uparrow} \nonumber \\
z_2&=&2z_2^{\uparrow}+\left(z_1^{\uparrow}\right)^2 \nonumber \\
z_3&=&2z_3^{\uparrow}+2z_2^{\uparrow}z_1^{\uparrow} \nonumber \\
z_4&=&2z_4^{\uparrow}+2z_3^{\uparrow}z_1^{\uparrow}+
\left(z_2^{\uparrow}\right)^2 \nonumber \\
z_5&=&2z_5^{\uparrow}+2z_4^{\uparrow}z_1^{\uparrow}+
2z_3^{\uparrow}z_2^{\uparrow} \nonumber \\
z_6&=&2z_6^{\uparrow}+2z_5^{\uparrow}z_1^{\uparrow}+
2z_4^{\uparrow}z_2^{\uparrow}+\left(z_3^{\uparrow}\right)^2 \nonumber \\
z_7&=&2z_7^{\uparrow}+2z_6^{\uparrow}z_1^{\uparrow}+
2z_5^{\uparrow}z_2^{\uparrow}+2z_4^{\uparrow}z_3^{\uparrow}.
\end{eqnarray}

The partition function for nuclei with protons and neutrons may now be 
constructed from the partition functions $z_n$. In Fig.~\ref{fig:fig1.ps} the 
relevant configurations for the lowest quasiparticle excitations in the 
proton-odd system are illustrated. For this one-quasiparticle system, there is only one proton partition function to 
consider. For the three-quasiparticle case the $\pi\nu^2$ and the $\pi^3$
partition functions have to be included. For the two-quasiparticle case of the 
even-even system we assume that either a $\pi$ or $\nu$ pair can be broken. In 
the four-quasiparticle case there are $\pi^2\nu^2$, $\pi^4$ 
and $\nu^4$ configurations to take into account. From similar arguments for 
higher number of quasiparticles, we construct the seven lowest partition 
functions for proton-odd and even-even systems by
\begin{eqnarray}
Z_1&=&z_1 \nonumber \\
Z_2&=&2z_2 \nonumber \\
Z_3&=&z_3+z_1z_2 \nonumber \\
Z_4&=&2z_4+(z_2)^2 \nonumber \\
Z_5&=&z_5+z_3z_2+z_1z_4 \nonumber \\
Z_6&=&2z_6+2z_4z_2 \nonumber \\
Z_7&=&z_7+z_5z_2+z_3z_4+z_1z_6.
\label{eq:Zeoee}
\end{eqnarray}
The partition functions for the neutron-odd system are identical to the 
functions for the proton-odd system. 

The partition functions for the odd-odd
system can be constructed in a similar manner. The two-quasiparticle partition function is simply 
$(z_1)^2$, giving an entropy of twice the single quasiparticle entropy of the 
odd-mass system. Higher numbers of quasiparticles yield the following partition
functions for the odd-odd case:
\begin{eqnarray}
\tilde{Z}_2&=&(z_1)^2 \nonumber \\
\tilde{Z}_4&=&2z_3z_1 \nonumber \\
\tilde{Z}_6&=&2z_5z_1+(z_3)^2 \nonumber \\
\tilde{Z}_8&=&2z_7z_1+2z_5z_3.
\end{eqnarray}

The single quasiparticle entropy is easily computed from our model with Eqs.~(1,2,\ref{eq:Zeoee}) as
\begin{equation}
S_{1}(T)=\frac{\epsilon/T}{\exp(\epsilon/T)-1}+
\ln\frac{2}{1-\exp(-\epsilon/T)}.
\end{equation}
At higher excitation energies more quasiparticles have to be considered and the
mathematical expressions become more complicated. In Fig.~\ref{fig:fig2.ps} the 
calculated $n$-quasiparticle entropy $S_n$ is shown as function of $\epsilon/T$. The entropy is seen to 
scale rather well with the number of quasiparticles, and there is a weak  
Pauli-blocking effect. The reason for this appearent extensivity is the onset of new degrees of freedom with 
increasing number of quasiparticles. In our formalism, this is manifested in 
the additional terms in the partition function appearing from the inclusion of 
spin up/down quasiparticles and the breaking of proton/neutron pairs.

This approximate extensivity of the entropy can be utilized to estimate the 
many quasiparticle partition function as a simple product of $n$ quasiparticle 
partition functions $Z_n=(Z_{1})^n$. The quasiparticle entropy is then 
roughly given by
\begin{equation}
S_n=nS_{1},
\end{equation}
where we argue in Sect.~IV that the $S_1$ value is experimentally available 
for most nuclei.

To create quasiparticles costs energy. In order to break one Cooper pair of the
core the energy $2\Delta$ is necessary. Taking this energy
into account we can finally write the partition functions of even-even (ee), odd (odd) and
odd-odd (oo) nuclei as weighted sums of the $n$-quasiparticle partition functions:
\begin{eqnarray}
Z^{\mathrm{ee}}&=&1+e^{-2\Delta/T}Z_2+e^{-4\Delta/T}Z_4+\ldots \nonumber \\
Z^{\mathrm{odd}}&=&Z_1+e^{-2\Delta/T}Z_3+e^{-4\Delta/T}Z_5+\ldots \nonumber \\
Z^{\mathrm{oo}}&=&\tilde{Z}_2+e^{-2\Delta/T}\tilde{Z}_4
+e^{-4\Delta/T}\tilde{Z}_6+\ldots
\label{eq:zoo}
\end{eqnarray}
provided an ideal core of Cooper pairs, not changing with the number of quasiparticles. If the system is extensive, we may write the partition functions in closed form. By using the property of extensivity, $Z_n=(Z_{1})^n$, and performing the sum up to 
$N$ broken Cooper pairs\footnote{For $N\rightarrow\infty$ broken Cooper pairs 
our simple model breaks down since the partition functions become infinite at 
the temperature $\epsilon/T=-\ln[1-2\,\exp(-\Delta/T)]$.}, we obtain
\begin{eqnarray}
Z^{\mathrm{ee}}&=&\frac{1-(e^{-\Delta/T}Z_1)^{2N+2}}
{1-(e^{-\Delta/T}Z_1)^2} \nonumber \\
Z^{\mathrm{odd}}&=&Z_1\,Z^{\mathrm{ee}} \nonumber \\
Z^{\mathrm{oo}}&=&\left(Z_1\right)^2\,Z^{\mathrm{ee}}.
\label{eq:zooex}
\end{eqnarray}
The latter partition functions are good approximations for the expressions of
Eq.~(\ref{eq:zoo}) for up to $\sim$2 broken pairs, however, for more than 
$\sim$4 quasiparticles, the Pauli blocking cannot be neglected.
 
Figure~\ref{fig:fig3.ps} shows the entropy as function of temperature 
calculated from our model, allowing up to 5 broken nucleon pairs. Also calculations assuming extensivity of the entropy, in the sense $Z_n =(Z_1)^n$, are shown. There are serious discrepances between the non-extensive and extensive approaches for $T > 0.3$ MeV. 

There are limitations of the present model. The excited states are 
assumed built by a few quasiparticles, while the other particles are forming
Cooper pairs in the underlying core. This picture will break down for 
({\em i}) light nuclei, ({\em ii}) around closed shells and 
({\em iii}) at high excitation energy. The first two points rest on the fact 
that the single particle energy spacing should be small or comparable to the 
pairing interaction in order to build a superfluid phase. The last point is 
connected to the thermal quenching of pair correlations, where many valence 
particles participate in the excitation. A fingerprint of the breakdown of the
thermal quasiparticle picture is when the entropy in the odd-mass and even-even
systems becomes essentially equal. These limitations have been addressed in 
Ref.~\cite{GB00}.

\section{Determination of level density anchor points}

The nuclear level density $\rho$ is scarcely known as function of excitation energy. Recently\cite{TD98}, there has been published compilations of level density parameters. In our work we show that it is possible to determine two level density points quite reliably for numerous nuclei. To demonstrate the technique, 
Fig.~\ref{fig:fig4.ps} shows the two extracted anchor points (filled data points) for $^{172}$Yb, together with the level density deduced from known discrete levels (solid lines). 
The first point $(E_1,\rho_1)$ is based on counting known discrete levels at low excitation 
energy, giving an anchor point in the excitation energy region of 
$0<E<2$~MeV. Information on known levels is taken from the database of 
Ref.~\cite{ENSDF}, where we have analyzed levels of $\sim 1600$ nuclei. The second point $(E_2,\rho_2)$ is estimated from the average neutron resonance spacing at the neutron binding energy $B_n$, where we use the compilation of Iljinov et al.\cite{Iljinov}.

Figure~\ref{fig:fig4.ps} also includes the data points\cite{Schi1} (open cirles) obtained with the method of Ref.\cite{Schi2}. The dashed straight line describes the data rather well. This line will be referred to as the constant-temperature level-density formula, defined by
\begin{equation}
\rho(E)=C e ^{E/T}
\label{eq:const}
\end{equation}
with $T^{-1}=(\ln \rho_2 - \ln \rho_1)/(E_2-E_1)$ and $C=\rho_1 \exp(-E_1/T)$. This line connecting the two anchor points is used to determine the nuclear temperature $T$.

In the extraction procedure of $(E_1,\rho_1)$, an initial slope is predicted for the straight line; in the case of $^{172}$Yb, the inverse of $T = 0.57$ MeV. Then the two highest values of the count-based $\ln \rho$ relative to this line is found\footnote{We pick the highest values since it is more probable that levels have escaped detection than the opposite.}. With maximums at the energy bin numbers $a$ and $b$, we find the lower anchor point by
\begin{equation}
\rho_1=\exp \left[ \frac{1}{b-a+1}\sum_{i=a}^b \ln \rho (i) \right]
\end{equation}
with energy $E_1$ in bin number $(a + b)/2$. In the evaluation of the uncertainty of $\rho_1$, the number of levels $N$ ($\sigma_N=\sqrt{N}$) and the uncertainty of $T$ ($\sigma_T=0.1T$) are taken into account. Nuclei far from $\beta$-stability may have been poorly investigated due to low reaction cross sections, and the extracted level density and uncertainties of these nuclei should be adopted with caution.

The second point $(E_2,\rho_2)$ is estimated from the average neutron resonance spacing at $B_n$.  Assuming neutron $\ell =0$ capture on a nucleus with target spin and parity $I^{\pi}$, levels with the same parity and two spins $I \pm 1/2$ are populated. In order to transfer the spacing $\overline{D}$ between these levels to the level density for all spins and parities, we use the spin distribution of Gilbert and Cameron \cite{GC}
\begin{equation}
\rho_2 = \frac{2 \sigma ^2}{\overline{D}} \left[ (I+1)e ^{\frac{-(I+1)^2}{2 \sigma ^2} } 
+ Ie ^{\frac{-I^2}{2 \sigma ^2 }} \right]^{-1}.
\end{equation}
The spin cut-off parameter $\sigma$ is defined through $\sigma^2 = 0.0888 A^{2/3}\sqrt{aU}$, where the back-shifted energy is $U=E-E_{bs}$. The level density parameter $a$ and the back-shift parameter $E_{bs}$ are defined by $a=0.21A^{0.87}$ MeV$^{-1}$ and $E_{bs} = C_{bs} + \Delta$, respectively, where the back-shift correction is given by $C_{bs}=-6.6A^{-0.32}$ according to Ref.~\cite{Egidy}. The pairing gap parameter is estimated by $\Delta = 12A^{-1/2}$ MeV.

In Figs.~\ref{fig:fig5.ps}-\ref{fig:fig7.ps} we show the anchor points extracted for nine isotopes of germanium, gadolinium and uranium, respectively. The above procedure has been tested carefully and found reliable, as demonstrated in these figures. In Table 1 the two anchor points are listed together with the temperature $T$ for the nuclei where both anchor points are known. We also see from Table 1 and Figs.~\ref{fig:fig5.ps}-\ref{fig:fig7.ps} that the extracted value of $T$ is  approximately equal for neighboring nuclei, thus, giving confidence to the methods applied for establishing the two anchor points.

\section{Extraction of entropy}

The measured level density $\rho(E)$ is proportional to the number of states accessible to the nuclear system at excitation energy $E$. Thus, the entropy in the microcanonical ensemble is given by 
\begin{equation}
S(E)=\ln (\rho (E)/\rho_0) = \ln \rho(E) + S_0.
\end{equation}
Applied to Eq.~(\ref{eq:const}), parameter $T$ is interpreted as the thermodynamical temperature in the microcanonical ensemble, since by definition:
\begin{equation}
\langle T(E) \rangle=\frac{1}{(\partial S/\partial E)_V}. 
\label{eq:t(e)}
\end{equation}

The constant shift parameter $S_0=-\ln \rho _0$ can be adjusted to fulfill the third law of thermodynamics; $S \rightarrow 0$ when $T \rightarrow 0$. The condition $T=0$ probably holds for the rotational ground band of deformed even-even nuclei. However, also other interpretations for $T=0$ are possible, and we will not fix the $S_0$ parameter to a specific value in this work, but instead discuss the relative entropy $S-S_0 = \ln \rho$.

The level density in between the two anchor points is not very well known. Recently\cite{Schi1,Melb1}, the level densities for $^{161,162}$Dy, $^{166}$Er and $^{171,172}$Yb have been measured. A common feature for these nuclei, is that the level densities develop more like the constant temperature level density formula of Eq.~(\ref{eq:const}), than the back-shifted Fermi-gas level density \cite{Egidy}, as pointed out in the  discussion of Fig.~\ref{fig:fig4.ps} and Ref.~\cite{GH00}. This constant temperature behavior is interpreted\cite{Schi1,Melb1} as the breaking of Cooper pairs with increasing excitation energy.

In the following, we will discuss the microcanonical entropy for given excitation energy $E$. In order to establish a common play ground, we extrapolate the first and second level density anchor points to excitation energies 1 and 7 MeV, using the constant temperature formula of Eq.~(\ref{eq:const}). Figures \ref{fig:fig8.ps} and \ref{fig:fig9.ps} display the entropy $S-S_0$ based on the two anchor points evaluated at 1 and 7 MeV, respectively. The data are plotted as function of the mass number, where nuclei along the $\beta$-stability line are chosen.

\section{Thermal quasiparticle properties}

\subsection{Regions of quasiparticle extensivity}

In the lower panel of figure \ref{fig:fig8.ps}  the difference in entropy between odd-mass and even-even systems $\Delta S = S_{\rm odd}-S_{\rm ee}$ is shown. The data show fluctuations which are partly due to shell closures, incomplete knowledge of discrete levels at low excitation energy, and structural changes between various isotopes and isotones. For the mid-shell regions the odd-mass nuclei have about $\Delta S \sim 1.7$ higher entropy compared to the even-even system. This feature holds for odd-even as well as for even-odd nuclei, however, neutrons carries slightly more entropy than protons. In Fig.~\ref{fig:fig9.ps} the data are sensitive to the level density extracted from neutron resonance level spacing. The fluctuations are less pronounced at 7 MeV because the level density is based on many quasiparticle states, smearing out structural effects. Also here, as indicated on the figure by lines, an entropy difference of $\Delta S \sim 1.7$ is found. Here, we should remember that these findings are based on a constant temperature level density extrapolation. If a Fermi gas level density is applied, e.g.~the model of Egidy et al.~shown in Fig.~\ref{fig:fig4.ps}, the extracted microcanonical entropy difference $\Delta S$ will be somewhat modified; it will increase at 1 MeV and decrease at 7 MeV. However, as demonstrated in Ref.~\cite{GB00}, the experimental value of $\Delta S$ is rather constant for excitation energies above 1.5 MeV.

It is amazing that $\Delta S$ is roughly the same both at 1 and 7 MeV of excitation energy. This means that even with 4 -- 6 quasiparticles excited at 7 MeV \cite{GB00}, the last thermal quasiparticle still carries $\Delta S \sim 1.7$. Thus, the experimental entropy scales with the number of quasiparticles. This property is also accounted for in our model\footnote{We should keep in mind that the model is formulated in the canonical ensemble, where a certain temperature $T$ corresponds to a wide excitation energy range.}. The regions of quasiparticle extensivity can be identified in Figs.~\ref{fig:fig8.ps} and \ref{fig:fig9.ps} as entropy-gaps between the entropies of odd-mass and even-even nuclei. The extensivity regions are found for mid-shell nuclei with mass number $A>40$.

In odd-odd nuclei, several neutron energy spacing data are known \cite{Iljinov}. As seen in Fig.~\ref{fig:fig9.ps} (see lines), a proton {\em and} a neutron give an entropy excess of around $2 \Delta S \sim 3.4$, as expected from the discussion in Sect.~II. A similar study from known discrete levels is difficult, due to experimental limitations in building level schemes with several hundreds of levels per MeV. However, in the case of the well studied odd-odd $^{152,154}$Eu isotopes the lower anchor point seems to have been determined with acceptable confidence. Figure \ref{fig:fig10.ps} shows that an additional neutron outside the odd-proton core gives an entropy difference as high as $2.5$. The extracted temperatures are all lying within a narrow window of $T=0.57 - 0.60$ MeV.

The feature that $\Delta S$ is approximately independent of excitation energy $E$ is shown in Table 1. Here, constant temperature $T$ for neighboring mid-shell nuclei means that the slope of $\ln \rho$ is equal and, thus, $\Delta S$ is approximately independent of $E$. Fig.~\ref{fig:fig11.ps} confirms the systematics, where neighboring nuclei exhibit the same $(\partial S/ \partial E)$ value. This indicates that the level density, and hence the entropy, has the same functional form as function of excitation energy. However, the feature disappears at shell closures.

\subsection{Constancy of quasiparticle entropy}

Figure \ref{fig:fig12.ps} shows the entropy difference for a neutron outside its even-even core for various rare earth elements. Although various mid-shell isotopes reveal quasiparticle entropy that deviates from $\Delta S \sim 1.7$ it is still surprising how constant this quantity appears to be. However, as shown in Fig.~\ref{fig:fig12.ps}, the quasiparticle around the $N=82$ shell closure carries low entropy due to the limited number of quantuum states (multiplicity)  available for the system.

\begin{figure}
\includegraphics[totalheight=17.5cm,angle=0,bb=0 80 350 730]{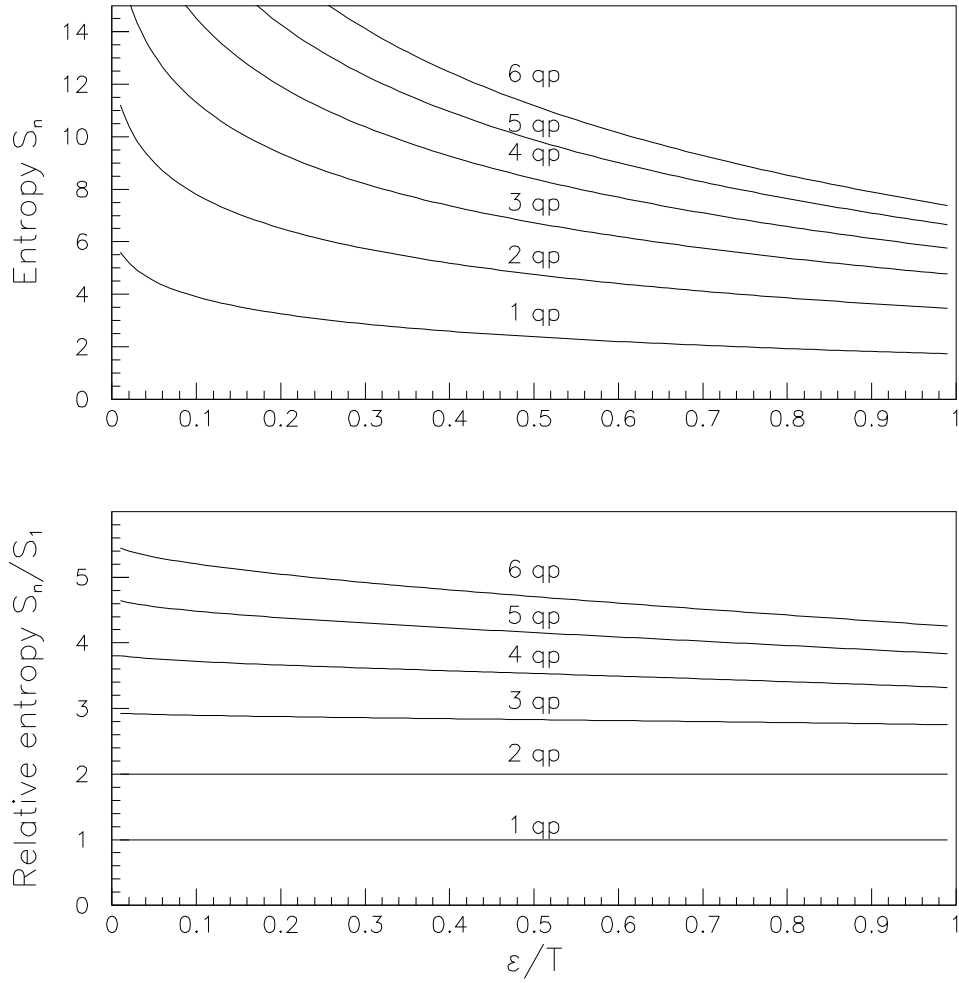}
\caption{The entropy $S_n$ of $n$ quasiparticles. The entropy increases with increasing temperature $T$ for fixed level spacing $\epsilon$. The lower part demonstrates that the Pauli blocking reduces the entropy per quasiparticle for $n>2$.}
\label{fig:fig2.ps}
\end{figure}

\begin{figure}
\includegraphics[totalheight=17.5cm,angle=0,bb=0 80 350 730]{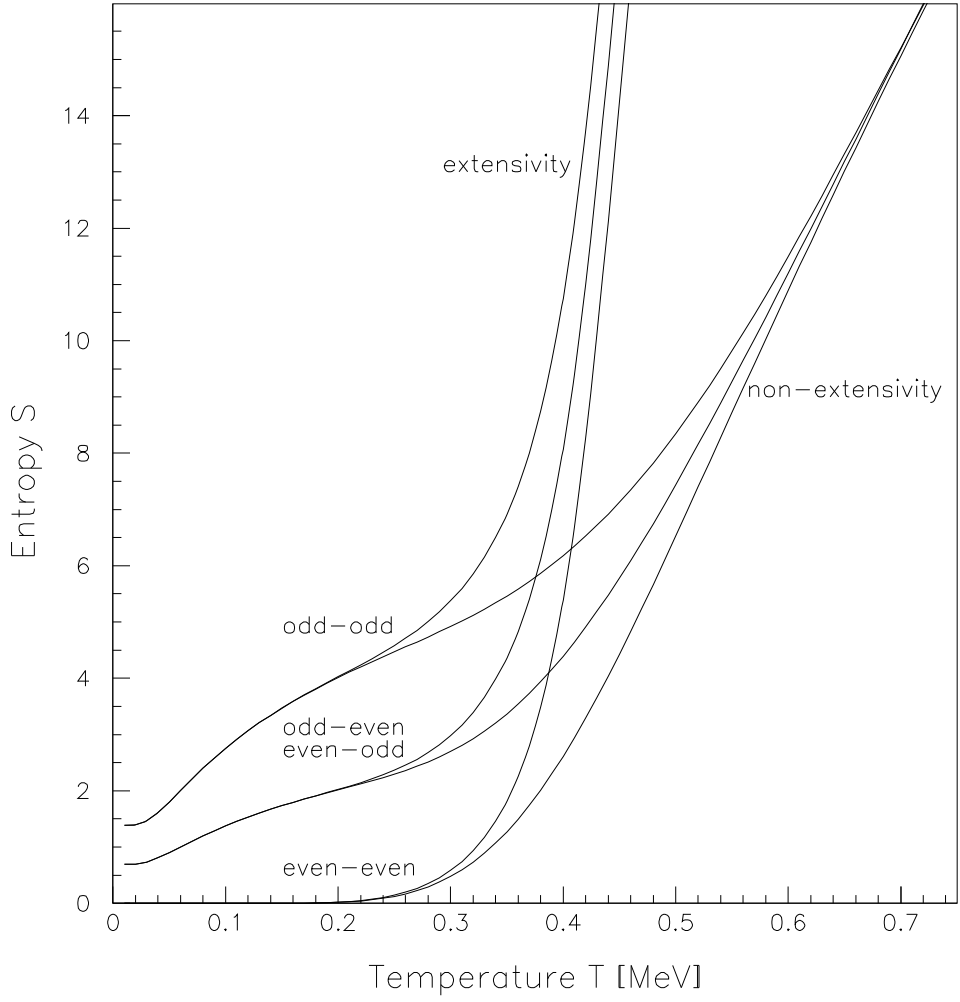}
\caption{The total entropy calculated for even-even, odd-mass and odd-odd nuclei allowing a maximum of five broken pairs. The extensive and non-extensive calculations are performed with Eqs.~(\ref{eq:zooex}) and (\ref{eq:zoo}), respectively. The parameters used are $\Delta$ = 0.9 MeV and $\epsilon$ = 0.15 MeV.}
\label{fig:fig3.ps}
\end{figure}

\begin{figure}
\includegraphics[totalheight=17.5cm,angle=0,bb=0 80 350 730]{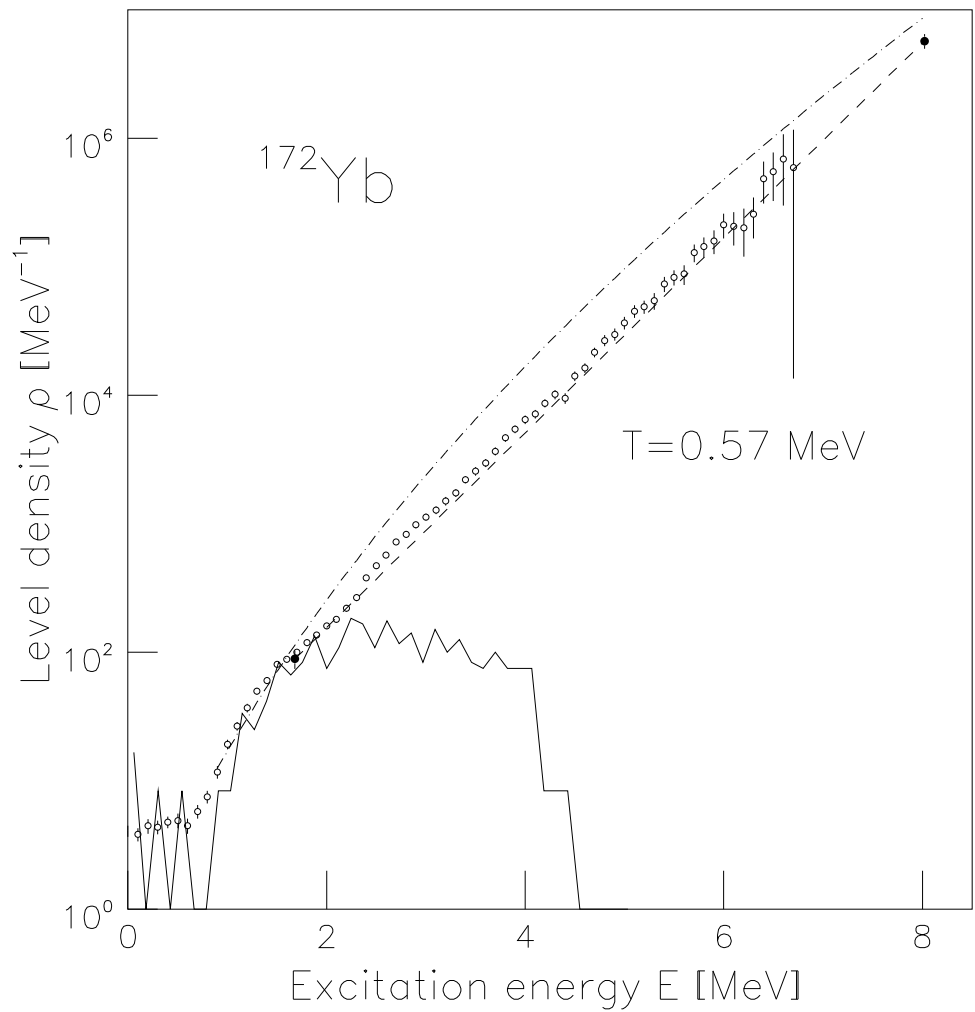}
\caption{Level densities in $^{172}$Yb, showing the two anchor points (filled circles). The solid line is based on the counting of discrete known levels. Also shown are the backshifted level density prediction of Egidy et al.~[9] (dashed-dotted) and the constant temperature formula (dashed).}
\label{fig:fig4.ps}
\end{figure}

\begin{figure}
\includegraphics[totalheight=17.5cm,angle=0,bb=0 80 350 730]{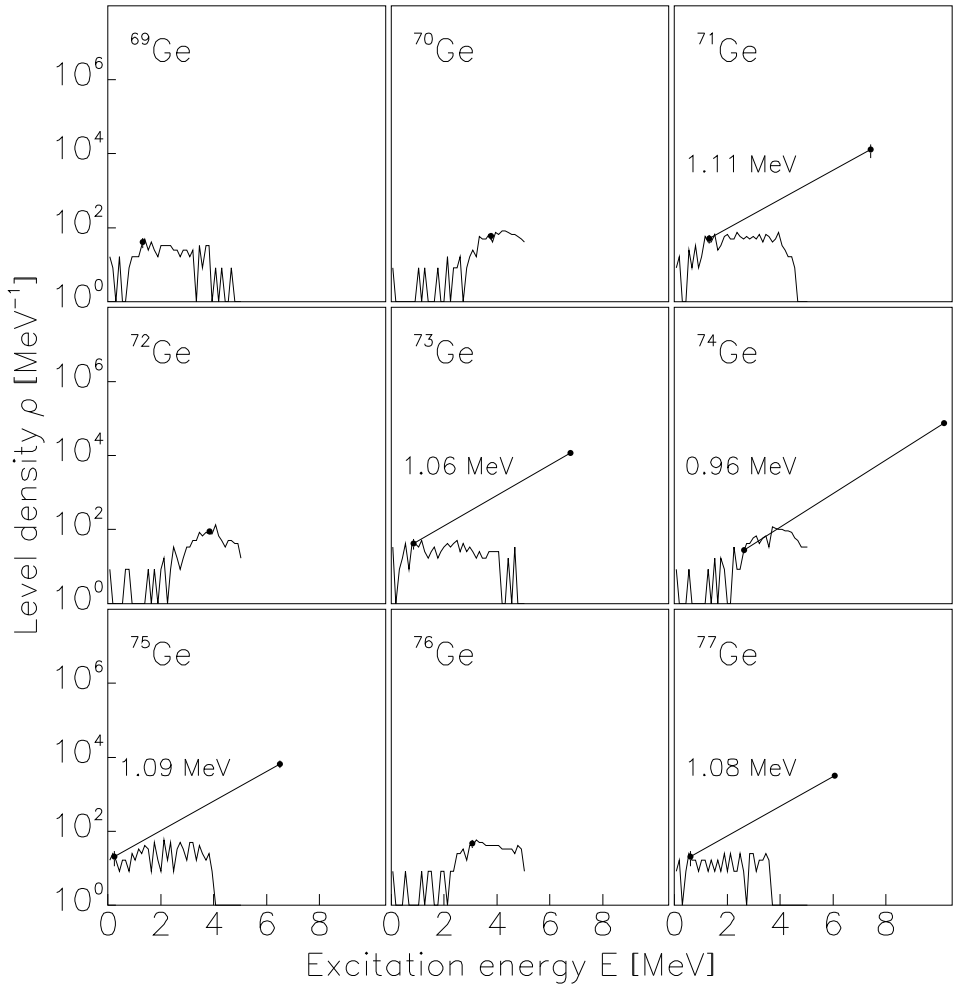}
\caption{The level density anchor points extracted for $^{69-77}$Ge. The lower point is based on known discrete levels. In cases where the second anchor point is known, the extracted temperature $T$ is quoted.}
\label{fig:fig5.ps}
\end{figure}

\begin{figure}
\includegraphics[totalheight=17.5cm,angle=0,bb=0 80 350 730]{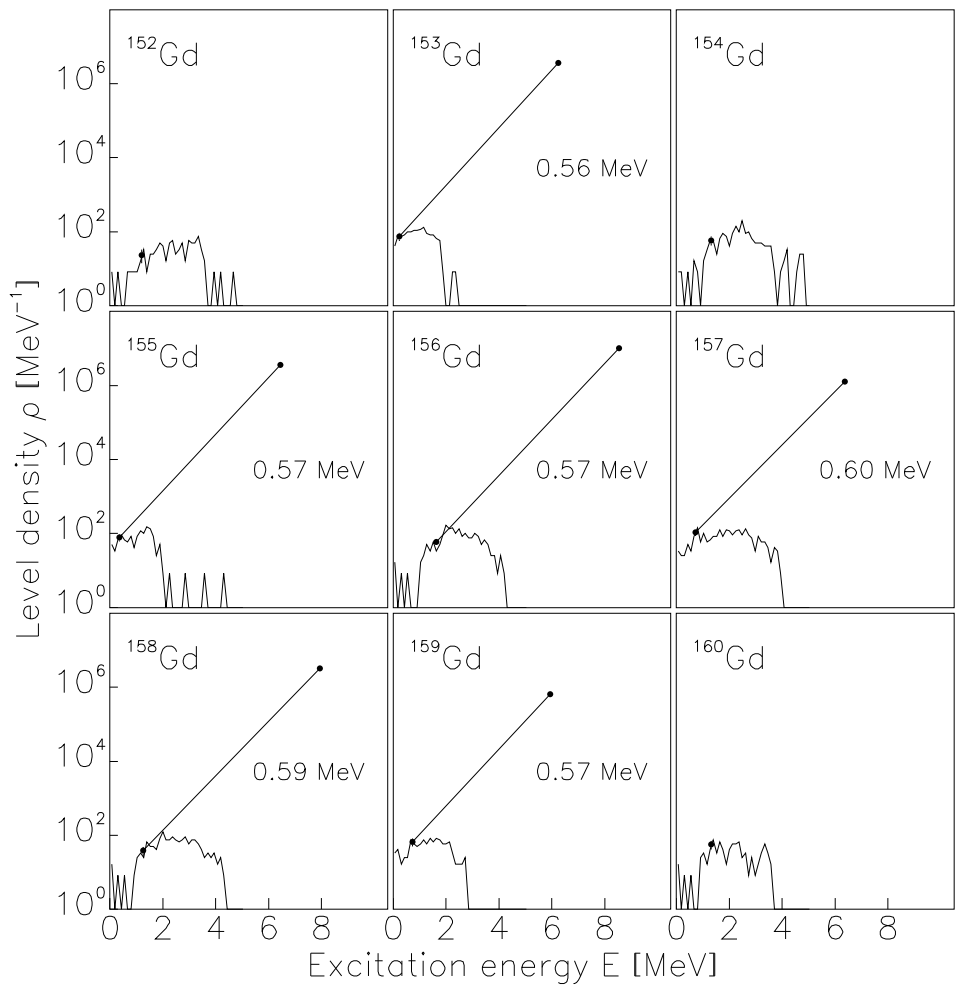}
\caption{Anchor points for $^{152-160}$Gd, see text of Fig.~\ref{fig:fig5.ps}.}
\label{fig:fig6.ps}
\end{figure}

\begin{figure}
\includegraphics[totalheight=17.5cm,angle=0,bb=0 80 350 730]{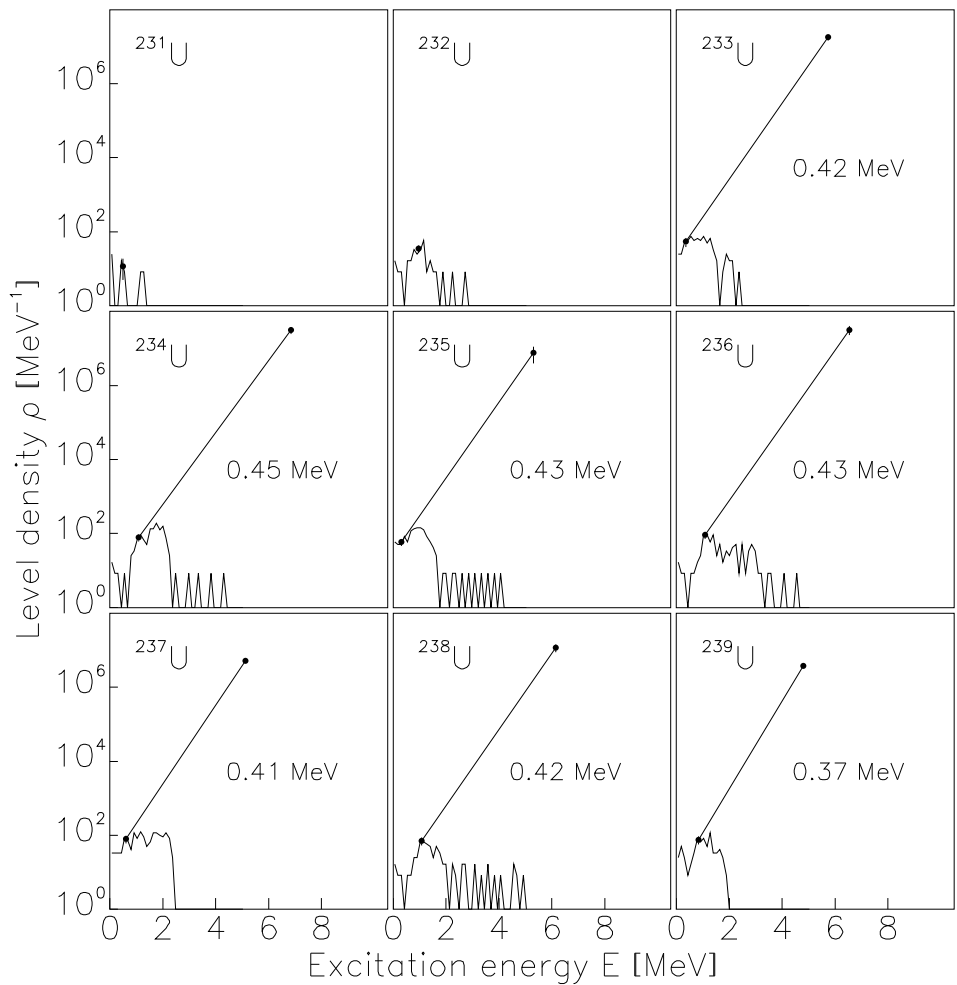}
\caption{Anchor points for $^{231-239}$U, see text of Fig.~\ref{fig:fig5.ps}.}
\label{fig:fig7.ps}
\end{figure}

\begin{figure}
\includegraphics[totalheight=17.5cm,angle=0,bb=0 80 350 730]{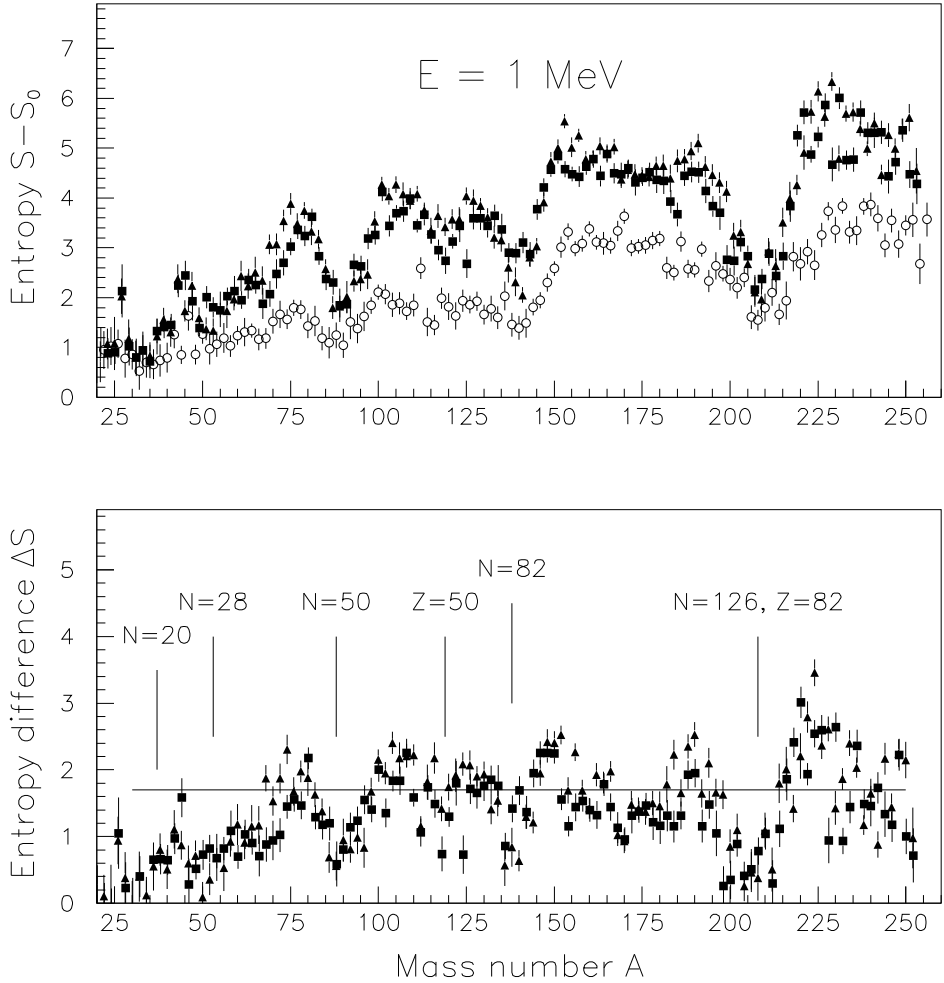}
\caption{Entropy as function of mass number at 1 MeV of excitation energy. The upper panel shows that the odd-even (filled triangles) and even-odd (filled squares) nuclei exhibit higher entropy than the even-even (open circles) nuclei. In the lower panel, the entropy difference between odd-mass and even-even nuclei is shown, giving an average of $\Delta S \sim 1.7$ (solid line) for mid-shell nuclei. The appoximate locations of the neutron and proton magic numbers are indicated. The data are based on the anchor points discussed in the text, and each data point is typically the average of 2-4 nuclei centered around the $\beta$-stability line.}
\label{fig:fig8.ps}
\end{figure}

\begin{figure}
\includegraphics[totalheight=17.5cm,angle=0,bb=0 80 350 730]{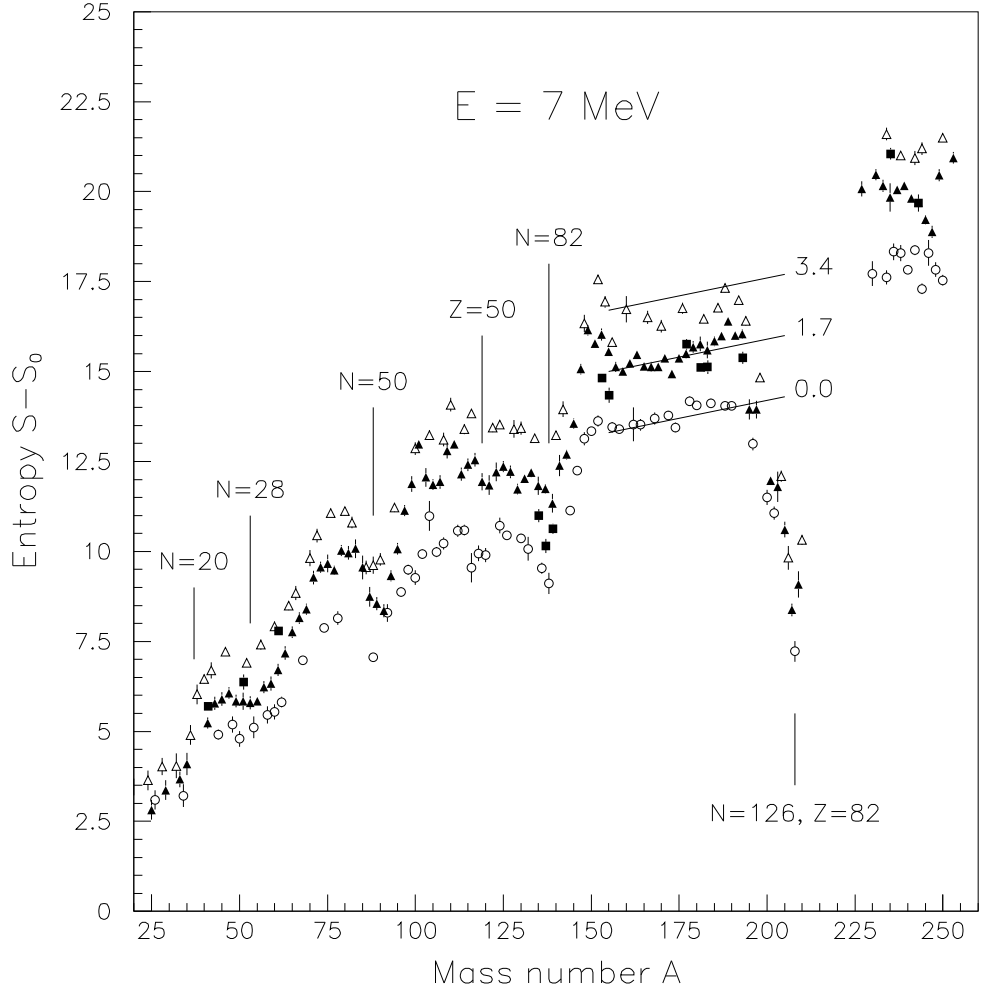}
\caption{Entropy as function of mass number at 7 MeV of excitation energy. The data are plotted for odd-odd (open triangles), odd-even (filled triangles), even-odd (filled squares) and even-even (open circles) nuclei, see text of Fig.~\ref{fig:fig8.ps}. The three lines are drawn to indicate the entropy gaps between even-even, odd-mass and odd-odd nuclei.}
\label{fig:fig9.ps}
\end{figure}

\begin{figure}
\includegraphics[totalheight=17.5cm,angle=0,bb=0 80 350 730]{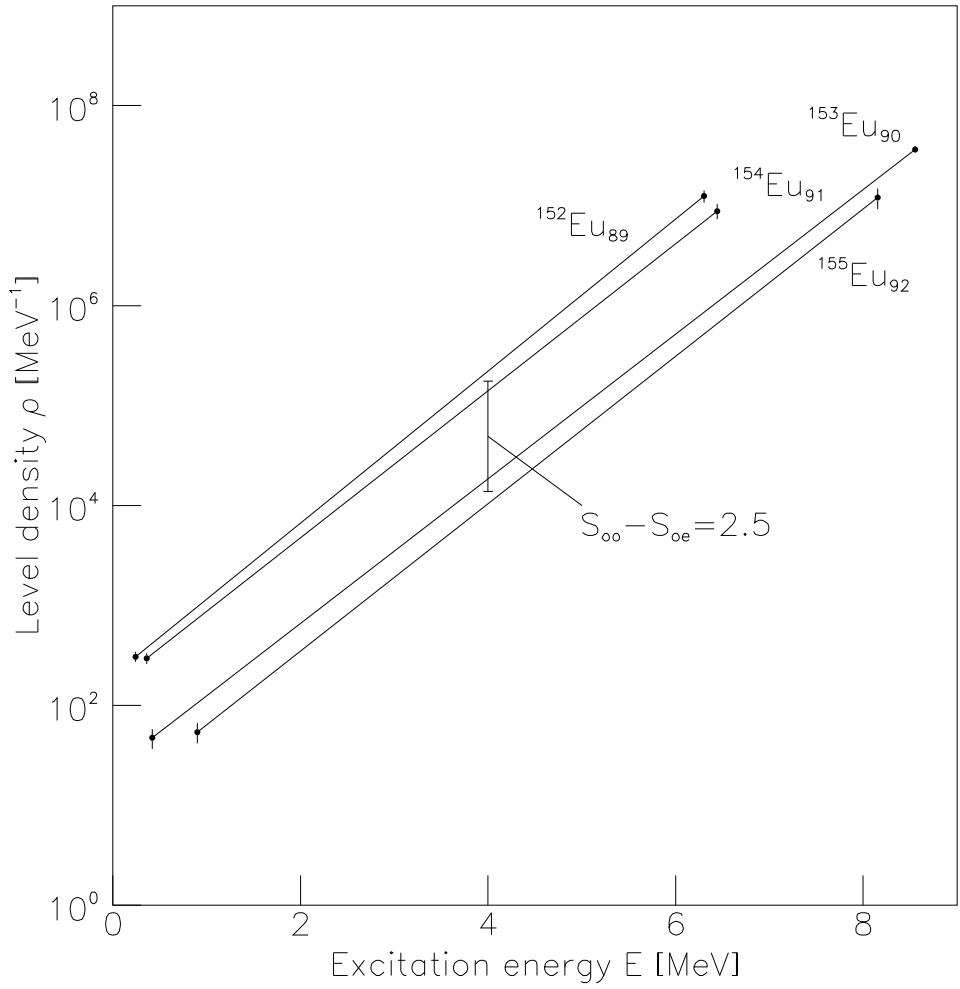}
\caption{Level densities and entropy difference for odd-even $^{153,155}$Eu and odd-odd $^{152,154}$Eu nuclei.}
\label{fig:fig10.ps}
\end{figure}

\begin{figure}
\includegraphics[totalheight=17.5cm,angle=0,bb=0 80 350 730]{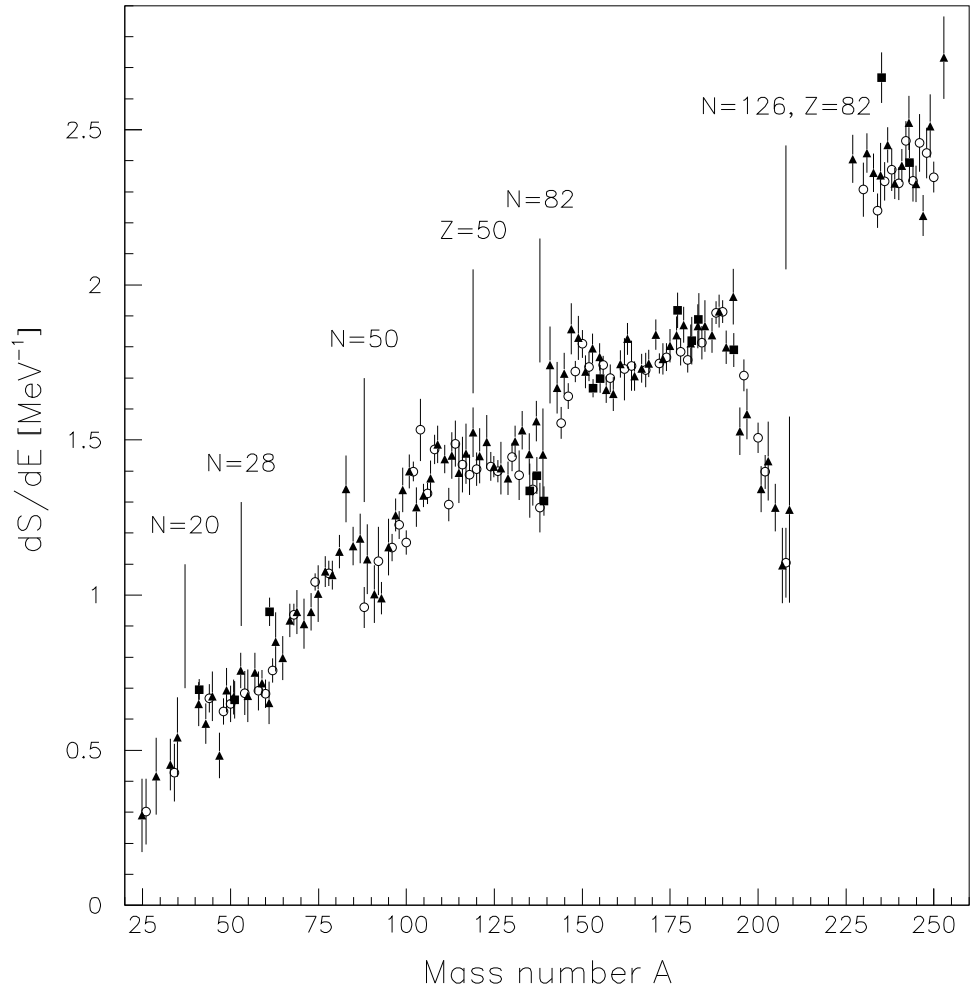}
\caption{The derivative of entropy with respect to excitation energy as function of mass number. The derivatives are determined from the anchor points of nuclei along the line of $\beta$-stability. The data are plotted for odd-even (filled triangles), even-odd (filled squares) and even-even (open circles) nuclei.}
\label{fig:fig11.ps}
\end{figure}

\begin{figure}
\includegraphics[totalheight=17.5cm,angle=0,bb=0 80 350 730]{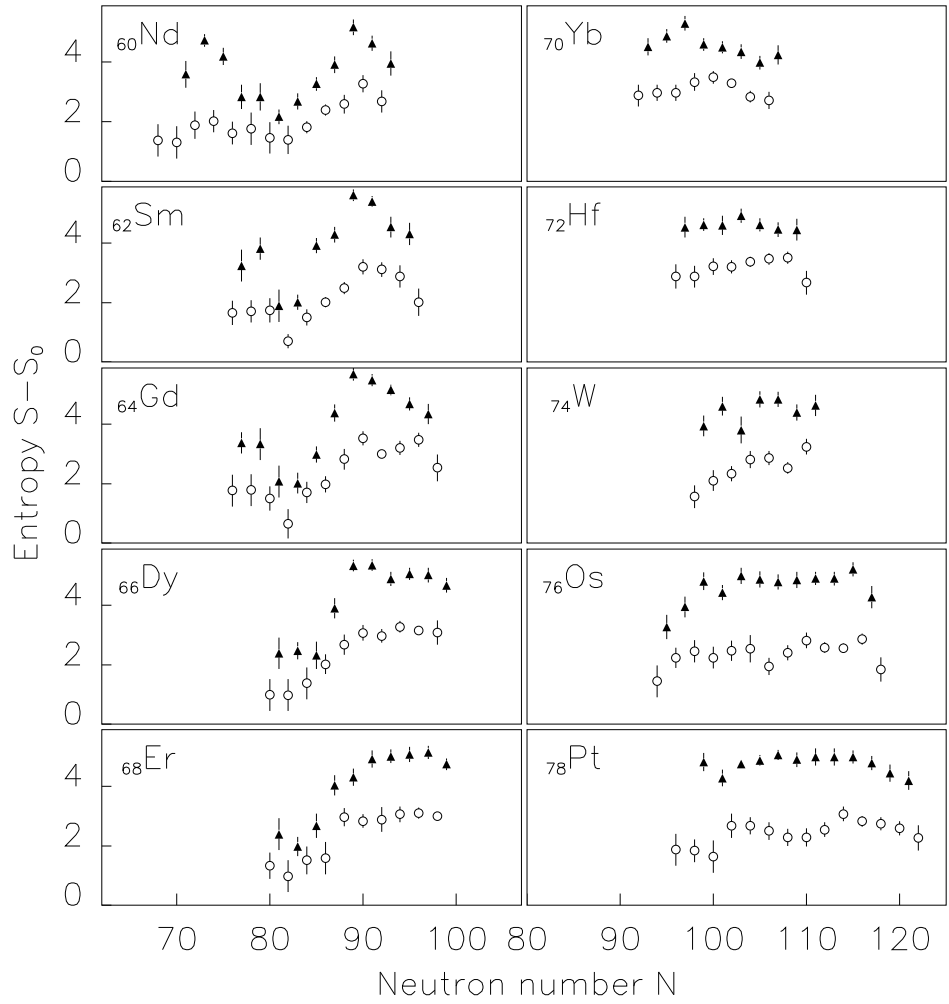}
\caption{The entropies extracted at 1 MeV of excitation energy for odd-even (filled triangles) and even-even (open cirles) rare earth isotopes. The data demonstrates that the entropy difference $\Delta S$ between odd-even and even-even isotopes is strongly reduced at the $N$ = 82 shell closure.}
\label{fig:fig12.ps}
\end{figure}

\begin{figure}
\includegraphics[totalheight=17.5cm,angle=0,bb=0 80 350 730]{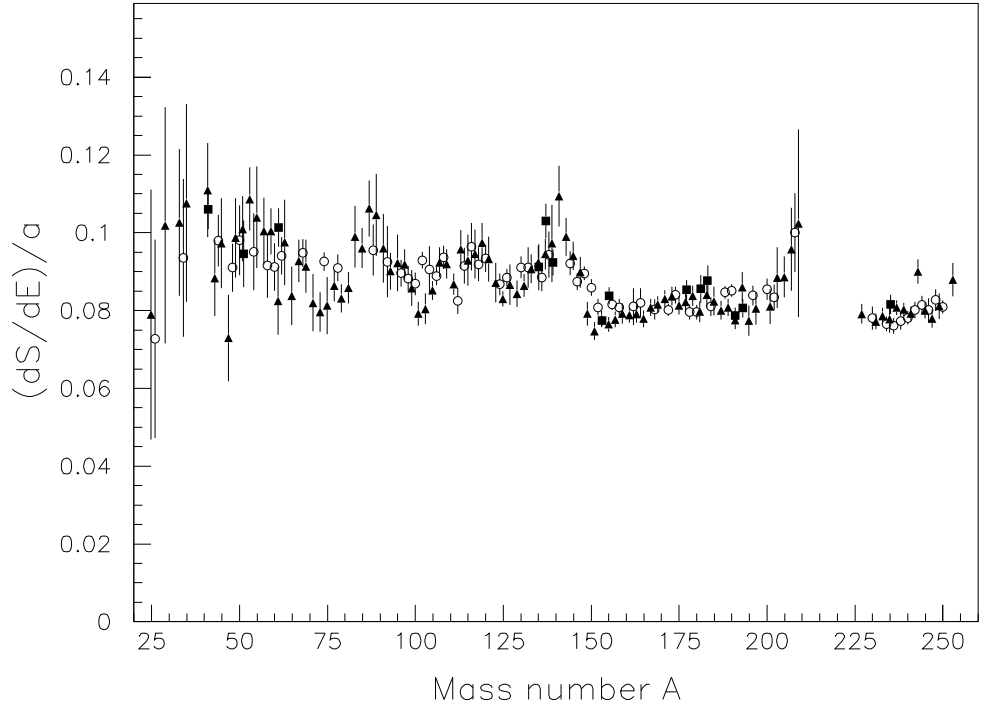}
\caption{The experimentally available quantity $(\partial S/ \partial E)/a$ is proportional to $\epsilon /T$, which determines the single quasiparticle entropy in our model.}
\label{fig:fig13.ps}
\end{figure}

Our model gives a qualitative interpretation of the constancy of the quasiparticle entropy. With the assumptions given, the model shows that the entropy $S_n$ is a function of $\epsilon /T$, only. In order to deduce experimentally the value of $\epsilon /T$, we recognize that the single particle level spacing $\epsilon$ is connected to the level density parameter $a$ by \cite{BM}
\begin{equation}
a=\frac{\pi^2}{6}(g_p + g_n) \sim \frac{\pi^2}{3}g=\frac{\pi^2}{3 \epsilon},
\label{eq:gpgn}
\end{equation}
where the single-particle level-density parameters for protons and neutrons ($g_p$ and $g_n$) are assumed to be approximately equal. Thus, we obtain using Eq.~(\ref{eq:t(e)})
\begin{eqnarray}
\frac{\epsilon}{T} =\frac{\pi^2}{3a}\left(\frac{\partial S}{\partial E}\right).
\label{eq:eps/t}
\end{eqnarray}
In Fig.~\ref{fig:fig13.ps} the quantity $(\partial S/ \partial E)/a$ is plotted as function of mass number $A$, using the $a$ values of Ref.~\cite{Iljinov} and the data of Fig.~\ref{fig:fig11.ps}. For increasing $A$, the extracted values are found in the range of $0.10 - 0.08$, which corresponds to $\epsilon /T = 0.33 - 0.26$ according to Eq.~(\ref{eq:eps/t}). This variation gives a minor effect on the predicted single particle entropy $S_1$; Fig.~\ref{fig:fig2.ps} shows that the variation in $\epsilon /T$ corresponds to an increase of $S_1$ from $2.7$ to $3.0$.

The experimentally extracted single quasiparticle entropy $\Delta S$ shows no clear systematic trend as function of $A$ within the uncertainties. Essentially, the local variations due to structural changes between neighboring nuclei are more pronounced than the weak variation in $S_1$ given by the model.

\begin{figure}
\includegraphics[totalheight=17.5cm,angle=0,bb=0 80 350 730]{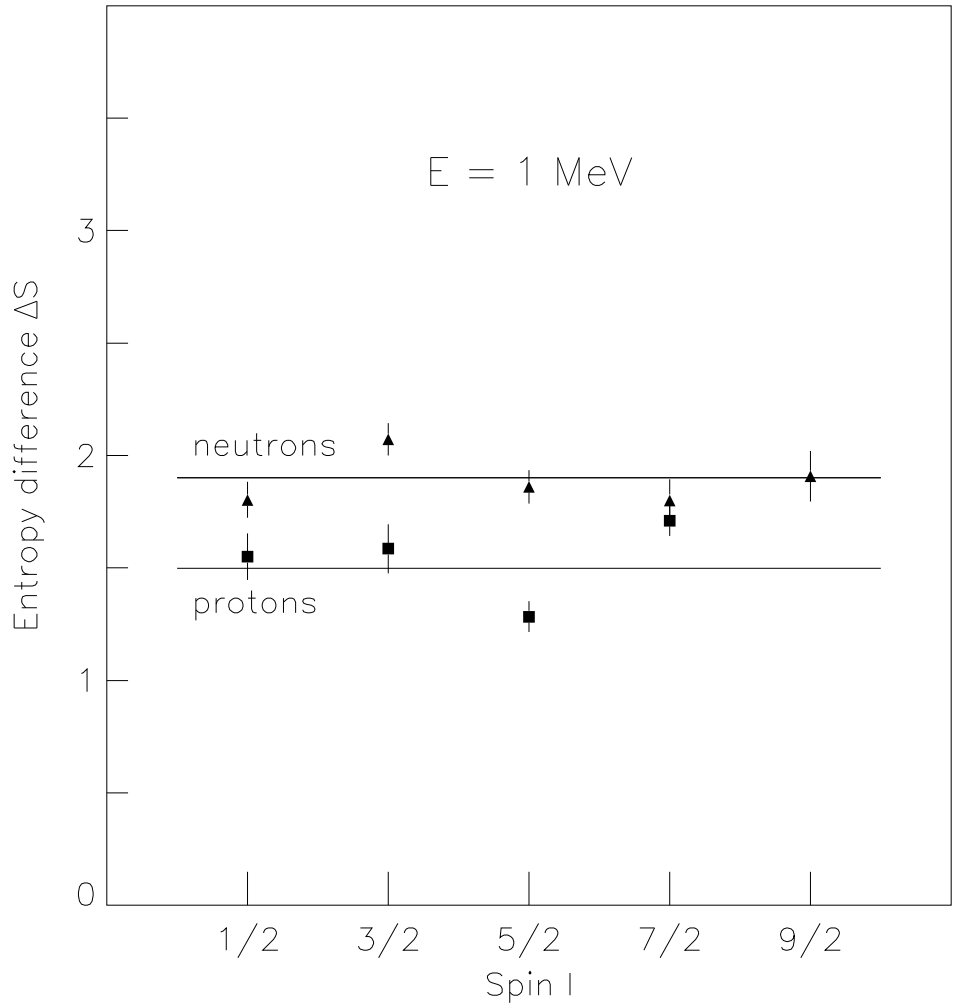}
\caption{The entropy difference $\Delta S$ for odd-even (filled triangles) and even-odd (filled squares) nuclei relative to even-even nuclei as function of ground state spin. The data are taken for nuclei with $Z=60-78$ and $N=90-110$. The solid lines are the averages for protons (1.5) and neutrons (1.9).}
\label{fig:fig14.ps}
\end{figure}

The quasiparticle entropy $\Delta S$ could in principle depend on the odd-mass ground state spin, since an orbital with high spin has more coupling combinations to the underlying rotating core. For excitation energies around the neutron binding energy $B_n$, the ground state orbital participates no more than other orbitals around the Fermi level, so no resemblence of the ground state spin is expected. However, around the lower anchor point ($E_1,\rho_1$), the ground state orbital could play a role. In Fig.~\ref{fig:fig14.ps} we see no significant effect for rare earth nuclei, and the same result applies to other mass regions. This indicates that many orbitals participate to build up the level density in the lower excitation region of odd-mass nuclei. 

To close this subsection, we point out that Fig.~\ref{fig:fig14.ps} indicates an average quasiparticle entropy $\Delta S$ of $\sim 1.5$ and $\sim 1.9$ for protons and neutrons, respectively. The valence protons and neutrons in this figure occupy orbitals in the $Z=50-82$ and $N=82-126$ shells, containing 32 and 44 orbitals, respectively. Thus, the single particle level energy spacing is expected to scale roughly as $\epsilon_p/\epsilon_n = 44/32 \sim 1.4$. According to Fig.~\ref{fig:fig2.ps}, a 40\% variation in $\epsilon/T$ around the value $\epsilon/T=0.3$ gives a change in $S_1$ of about 0.4. Therefore, this effect could be the reason why the protons carry slightly less entropy compared to the neutrons.

\section{Application of thermal quasiparticles}

There are several applications for the concepts presented here. In analogy with the description of rotational nuclei\cite{Frau}, the Helmholtz free energy $F$ plays the same role as the energy in the rotating frame $E'$. To visualize the analogy, the two cases read
\begin{eqnarray}
F&=&E-TS \\
\label{eq:ets}
E'&=&E-\omega I_x,
\end{eqnarray}
where $\omega$ is the rotational frequency and $I_x$ is the spin alignment along the rotational x-axis. Generally, $S$ and $I_x$ are positive quantities, so that $F$ and $E'$ decreases with increasing $T$ and $\omega$, respectively. A decoupled Cooper pair with high $I_x$ may, at a certain critical frequency $\omega_c$, become more favoured than its coupled counterpart, giving rise to the rotational backbending phenomenon. 

In a similar manner, we will here demonstrate that the free energy of a system with two additional excited quasiparticles becomes favoured at a critical temperature $T_c$. This temperature depends on the pairing gap $\Delta$ and the quasiparticle entropy $\Delta S$. If both these quantities would be independent of the number of quasiparticles excited, a break-up of all Cooper pairs should happen at one and the same critical temperature $T_c$, giving rise to a strong quenching of the pair correlations. 

Let us assume that the partition function for an even-even nucleus is described by $Z(T)=Z_{\rm core}(T)$. This partition function represents the lowest free energy $F(T)$ at low temperatures $T$. We now introduce the breaking of one pair relative to this core\footnote{The core described by $Z_{\rm core}$ is not necessary a cold core, but may already have pairs broken.}. Assuming the two quasiparticles to be independent of the core, we may describe the excited system by a factorization of the partition function into
\begin{equation}
Z^* \approx Z_{\rm core}Z_2 e^{-2\Delta /T},
\end{equation}
where $Z_2$ is the two quasiparticle partition function. With this assumption, the difference in Helmholtz free energy becomes
\begin{equation}
\Delta F=-T(\ln Z^* - \ln Z_{\rm core})=-T \ln Z_2 +2\Delta,
\end{equation}
and factorizing $Z_2$ assuming extensivity, gives
\begin{equation}
\Delta F= -T \ln Z_1 ^2 + 2\Delta=2(F_1 + \Delta),
\label{eq:f1}
\end{equation}
where $F_1$ is the free energy of one quasiparticle. This corresponds to the chemical potential defined (see Eq.~(\ref{eq:muT})) as the energy required to bring one quasiparticle out of the reservoir:
\begin{equation}
\mu(T)=\Delta F /2=F_1 + \Delta.
\end{equation}

\begin{figure}
\includegraphics[totalheight=17.5cm,angle=0,bb=0 80 350 730]{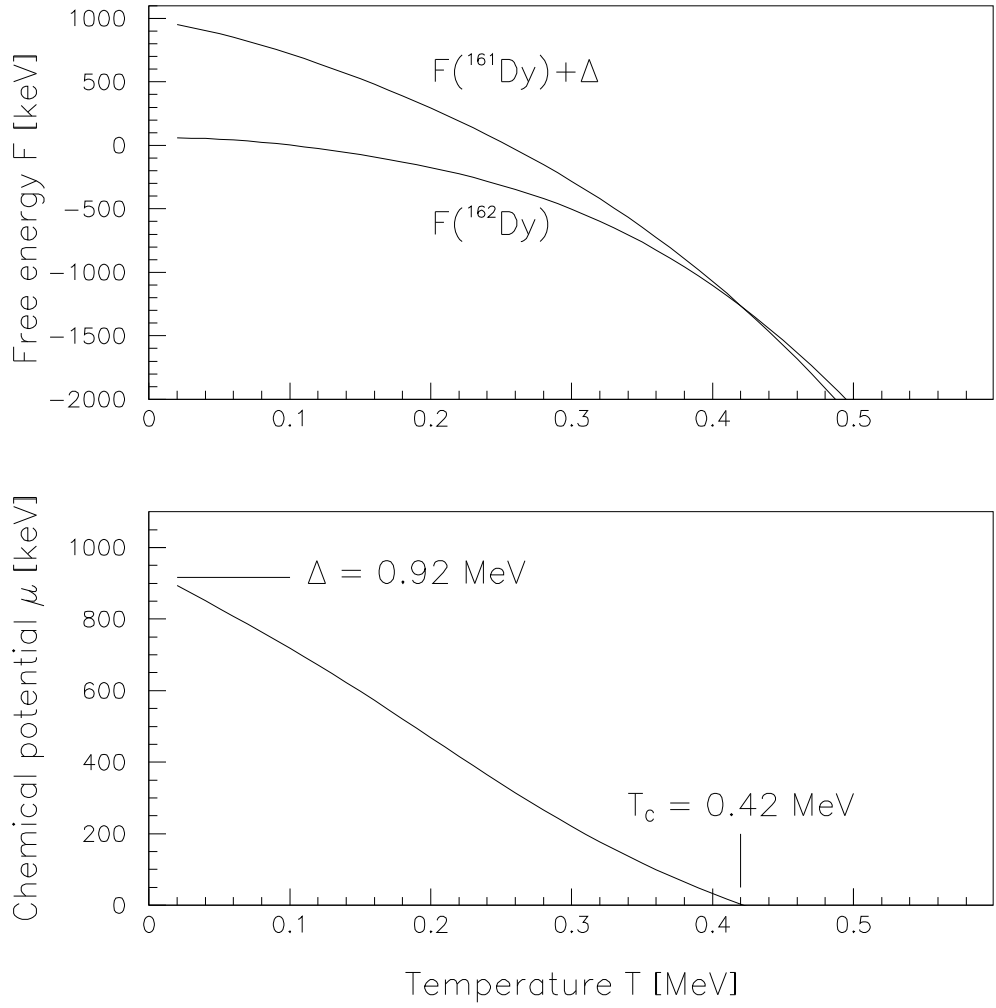}
\caption{The experimentally deduced Helmholtz free energy for $^{161,162}$Dy extracted in the canonical ensemble. The critical temperature for the quenching of pair correlations is found at $T_c = 0.42$ MeV, where the chemical potential is $\mu \sim 0$.}
\label{fig:fig15.ps}
\end{figure}

We have applied this formalism to the experimental data of $^{161,162}$Dy, see Ref.~\cite{GB00}. The partition functions for each nucleus are calculated by
\begin{equation}
         Z(T)=\sum_E \Delta\!E\rho(E)e^{-E/T}
         \label{eq:zactual}
\end{equation}
where $\Delta\!E$ is the experimental energy bin. The assumptions made for evaluating $Z$ is given elsewhere\cite{GB00}. The corresponding Helmholtz free energy $F$ for $^{162}$Dy is displayed in the upper panel of Fig.~\ref{fig:fig15.ps}. The upper energy curve, based on data for $^{161}$Dy, is given by $F(^{161}$Dy) + $\Delta_n$, where the pairing gap parameter is determined from neutron separation energies ${\cal S}_n$ by
\begin{equation}
\Delta_n  =  \frac{1}{4}|{\cal S}_n (N+1,Z) - 2{\cal S}_n (N,Z) + {\cal S}_n (N-1,Z)|,
\end{equation}
giving $\Delta_n =$ 0.917 MeV for $^{162}$Dy. Thus, the energy curve of $^{161}$Dy can be interpreted as the free energy for an even-even system with one extra quasiparticle. We therefore identify $F_1 + \Delta_n$ in Eq.~(\ref{eq:f1}) with the difference $F(^{161}{\rm Dy})+ \Delta_n - F(^{162}{\rm Dy})$. 

In the lower panel of Fig.~\ref{fig:fig15.ps} the chemical potential $\mu=F_1 +\Delta_n$ is shown as function of temperature. The higher entropy in $^{161}$Dy compared to $^{162}$Dy decreases the free energy relative to the even-even core with increasing temperature. When $\mu(T) \sim 0$, thermal quasiparticles can be formed without consuming free energy. Thus, from Fig.~\ref{fig:fig15.ps} we estimate the critical temperature\footnote{At the critical temperature $T_c$ we have $F_1=-\Delta=-T_c\ln Z_1$, and applying $Z_1=2z_1^{\uparrow}$, we get $\epsilon/T_c=-\ln[1-2\,\exp(-\Delta/T_c)]$, the same expression as in footnote 1.} for the quenching of pair correlations to be $T_c = 0.42$ MeV. The uncertainty of this number is mainly due to the extensivity assumed in the evaluation of $\Delta F$. Furthermore, the adoption of a chemical potential $\mu = \Delta$ at $T=0$, rests on the assumption that the Fermi level coincides with single particle energies in both nuclei. Also the extrapolation of the experimental level density to higher excitation energies, see Ref.~\cite{Schi1}, gives systematical errors. From these considerations we estimate a 15 \% error in the $T_c$ value.

The deduced value of $T_c=0.42(6)$ is considered to be common to the two dysprosium isotopes analyzed. Recently\cite{Schi1}, values of $T_c =$ 0.52(4) and 0.49(4) MeV were found for $^{161}$Dy and $^{162}$Dy, respectively, using the canonical heat capacity as thermometer for the phase transition.

Finally, we point out a curious connection between the experimental quasiparticle entropy $\Delta S$ and the critical temperature $T_c$ deduced from a Fermi gas model with pairing. If we assume a constant $\Delta S$ and a constant differences $\Delta$ between the $E$ functions for the two curves of  Fig.~\ref{fig:fig15.ps}, we find 
\begin{equation}
T_c=\frac{1}{\Delta S}\Delta,
\end{equation} 
giving $T_c=0.59\Delta$ for $\Delta S=1.7$. This estimate coincides almost exactly with the theoretical expression $T_c=0.57 \Delta$ given in Refs.~\cite{SY63,IG79}. The theoretical factor $0.57$ is independent on the mass number, consistent with the property of the experimental single quasiparticle entropy.

\section{Conclusions}

We have analyzed the level densities of nuclei based on counting discrete levels and from neutron scattering data. It was possible to obtain level density anchor points for about 280 nuclei. The level densities have been transformed into microcanonical entropy, revealing the phase space of the system at fixed excitation energies. 

The single quasiparticle entropy has been extracted from the odd-even entropy differences. The quasiparticles not coupled in Cooper pairs, exhibit an entropy of $\Delta S \sim 1.7$ which is relatively independent of the presence of other quasiparticles. This apparent extensivity is valid for mid-shell nuclei and for all ground state spins. A simple model for thermal quasiparticles in the canonical ensemble also supports the property of extensivity. 

The single quasiparticle entropy $\Delta S$ for mid-shell nuclei shows no clear mass dependency. This is explained in our model by the fact that $\Delta S$ only depends on the ratio between the single particle spacing $\epsilon$ and the temperature $T$, which is found to be almost independent of the mass number for mid-shell nuclei.

For thermal systems, the Helmholtz free energy is analogous to the energy in the internal frame of rotating nuclei. This opens for the study of various quantities and critical phenomena as function of temperature. We therefore believe that the concept of single quasiparticle entropy may provide a fruitful basis for the understanding of hot nuclei.

The simple model presented here is only adequate for mid-shell nuclei. Around closed shells one expects exciting effects from the increasing single particle energy spacings. This will also influence the entropy difference between odd-mass and even-even nuclei. Therefore, a statistical description of the transition to closed shells would be of great interest.

We wish to acknowledge the support from the Norwegian Research Council (NFR).

\end{multicols}

\begin{table}
Table 1: Level density anchor points and temperatures.\\
\begin{tabular}{llllll}
Nuclide   &$E_1$(MeV)& $\rho_1$(MeV$^{-1}$) & $E_2$(MeV) & $\rho_2$ (MeV$^{-1}$) & $T$(MeV)\\ \hline
$^{20}$F  &     2.70 &    6( 2) &     6.60 & 0.27E+02(0.70E+01) &  2.53(63) \\
$^{24}$Na &     1.62 &    6( 3) &     6.96 & 0.38E+02(0.12E+02) &  3.04(90) \\
$^{25}$Mg &     0.78 &    3( 2) &     7.33 & 0.18E+02(0.57E+01) &   3.4(14) \\
$^{26}$Mg &     4.92 &   12( 7) &    11.09 & 0.76E+02(0.24E+02) &   3.3(12) \\
$^{28}$Al &     1.32 &    5( 2) &     7.73 & 0.77E+02(0.21E+02) &  2.29(43) \\
$^{29}$Si &     2.22 &    4( 3) &     8.47 & 0.54E+02(0.17E+02) &  2.40(71) \\
$^{32}$P  &     3.36 &    9( 2) &     7.94 & 0.91E+02(0.37E+02) &  2.00(43) \\
$^{33}$S  &     2.46 &    5( 2) &     8.64 & 0.82E+02(0.21E+02) &  2.20(41) \\
$^{34}$S  &     2.70 &    4( 3) &    11.42 & 0.16E+03(0.62E+02) &  2.34(51) \\
$^{35}$S  &     3.06 &    7( 2) &     6.99 & 0.60E+02(0.22E+02) &  1.84(44) \\
$^{36}$Cl &     3.24 &   12( 2) &     8.58 & 0.37E+03(0.12E+03) &  1.56(17) \\
$^{38}$Cl &     1.86 &   10( 5) &     6.11 & 0.22E+03(0.69E+02) &  1.39(26) \\
$^{41}$Ar &     0.36 &    3( 2) &     6.10 & 0.14E+03(0.37E+02) &  1.52(31) \\
$^{40}$K  &     2.58 &   21( 6) &     7.80 & 0.12E+04(0.16E+03) &  1.30(10) \\
$^{41}$K  &     3.90 &   34( 6) &    10.10 & 0.26E+04(0.26E+03) &  1.44( 7) \\
$^{42}$K  &     1.44 &   37(12) &     7.53 & 0.11E+04(0.28E+03) &  1.81(23) \\
$^{41}$Ca &     3.12 &   11( 3) &     8.36 & 0.34E+03(0.82E+02) &  1.54(17) \\
$^{43}$Ca &     2.88 &   29( 7) &     7.93 & 0.56E+03(0.11E+03) &  1.71(19) \\
$^{44}$Ca &     3.90 &   17( 5) &    11.13 & 0.21E+04(0.33E+03) &  1.50(10) \\
$^{45}$Ca &     1.68 &   10( 4) &     7.42 & 0.48E+03(0.98E+02) &  1.49(18) \\
$^{49}$Ca &     3.00 &    9( 3) &     5.15 & 0.57E+02(0.20E+02) &  1.14(33) \\
$^{46}$Sc &     0.24 &   17( 9) &     8.76 & 0.43E+04(0.33E+03) &  1.54(16) \\
$^{47}$Ti &     2.70 &   54(23) &     8.88 & 0.11E+04(0.19E+03) &  2.07(32) \\
$^{48}$Ti &     4.02 &   28( 6) &    11.63 & 0.32E+04(0.82E+03) &  1.60(11) \\
$^{49}$Ti &     2.58 &   19( 7) &     8.14 & 0.91E+03(0.17E+03) &  1.44(15) \\
$^{50}$Ti &     3.42 &   12( 4) &    10.94 & 0.16E+04(0.39E+03) &  1.54(14) \\
$^{51}$Ti &     1.92 &    6( 2) &     6.37 & 0.24E+03(0.14E+03) &  1.23(23) \\
$^{51}$V  &     3.54 &   59(23) &    11.05 & 0.86E+04(0.20E+04) &  1.51(14) \\
$^{52}$V  &     1.32 &    9( 3) &     7.31 & 0.13E+04(0.16E+03) &  1.22( 8) \\
$^{51}$Cr &     1.14 &    6( 2) &     9.26 & 0.13E+04(0.22E+03) &  1.49(12) \\
$^{53}$Cr &     1.86 &    7( 2) &     7.94 & 0.67E+03(0.13E+03) &  1.32(10) \\
$^{54}$Cr &     2.64 &    8( 3) &     9.72 & 0.11E+04(0.38E+03) &  1.46(15) \\
$^{55}$Cr &     0.42 &    6( 3) &     6.25 & 0.30E+03(0.54E+02) &  1.47(23) \\
$^{56}$Mn &     0.66 &   15( 5) &     7.27 & 0.20E+04(0.32E+03) &  1.35(11) \\
$^{55}$Fe &     0.66 &    3( 2) &     9.30 & 0.11E+04(0.22E+03) &  1.48(19) \\
$^{57}$Fe &     0.66 &    4( 2) &     7.65 & 0.83E+03(0.15E+03) &  1.33(11) \\
$^{58}$Fe &     2.46 &   10( 4) &    10.05 & 0.19E+04(0.52E+03) &  1.45(13) \\
$^{59}$Fe &     0.84 &    8( 4) &     6.58 & 0.57E+03(0.25E+03) &  1.36(20) \\
$^{60}$Co &     1.08 &   19( 3) &     7.49 & 0.42E+04(0.61E+03) &  1.19( 5) \\
$^{61}$Co &     1.44 &   12( 4) &     9.32 & 0.22E+05(0.16E+04) &  1.06( 5) \\
$^{59}$Ni &     1.02 &    6( 2) &     9.00 & 0.17E+04(0.29E+03) &  1.40( 8) \\
$^{60}$Ni &     3.84 &   30( 8) &    11.40 & 0.51E+04(0.12E+04) &  1.47(10) \\
$^{61}$Ni &     2.04 &   32(11) &     7.82 & 0.14E+04(0.26E+03) &  1.53(16) \\
$^{62}$Ni &     3.66 &   27( 6) &    10.60 & 0.51E+04(0.84E+03) &  1.32( 7) \\
$^{63}$Ni &     1.14 &    9( 5) &     6.84 & 0.11E+04(0.24E+03) &  1.18(13) \\
$^{65}$Ni &     0.48 &    4( 3) &     6.10 & 0.75E+03(0.17E+03) &  1.05(15) \\
$^{64}$Cu &     1.68 &   49( 7) &     7.92 & 0.11E+05(0.14E+04) &  1.16( 4) \\
$^{66}$Cu &     0.30 &   13( 6) &     7.07 & 0.74E+04(0.16E+04) &  1.07( 8) \\
$^{65}$Zn &     0.96 &   25(10) &     7.98 & 0.67E+04(0.20E+04) &  1.26(11) \\
$^{67}$Zn &     1.86 &   31( 7) &     7.05 & 0.36E+04(0.66E+03) &  1.09( 6) \\
$^{68}$Zn &     3.42 &   37( 8) &    10.20 & 0.21E+05(0.26E+04) &  1.07( 4) \\
$^{69}$Zn &     1.20 &   18( 6) &     6.48 & 0.27E+04(0.46E+03) &  1.06( 8) \\
$^{71}$Zn &     0.60 &    8( 3) &     5.83 & 0.33E+04(0.58E+03) &  0.86( 7) \\
$^{70}$Ga &     1.14 &   57(12) &     7.65 & 0.35E+05(0.88E+04) &  1.02( 5) \\
$^{72}$Ga &     0.54 &   44( 7) &     6.52 & 0.21E+05(0.43E+04) &  0.97( 4) \\
$^{71}$Ge &     1.32 &   51(14) &     7.42 & 0.13E+05(0.53E+04) &  1.10(10) \\
$^{73}$Ge &     0.84 &   42(13) &     6.78 & 0.12E+05(0.20E+04) &  1.06( 7) \\
$^{74}$Ge &     2.64 &   28( 5) &    10.20 & 0.74E+05(0.80E+04) &  0.96( 3) \\
$^{75}$Ge &     0.24 &   20( 9) &     6.51 & 0.66E+04(0.14E+04) &  1.09( 9) \\
$^{77}$Ge &     0.60 &   20( 9) &     6.07 & 0.32E+04(0.44E+03) &  1.08(10) \\
$^{76}$As &     0.54 &   82(15) &     7.33 & 0.89E+05(0.12E+05) &  0.97( 3) \\
$^{75}$Se &     1.08 &   62(16) &     8.03 & 0.67E+05(0.39E+05) &  1.00( 9) \\
$^{77}$Se &     0.96 &   34( 8) &     7.42 & 0.36E+05(0.80E+04) &  0.93( 4) \\
$^{78}$Se &     2.52 &   29( 7) &    10.50 & 0.15E+06(0.30E+05) &  0.94( 4) \\
$^{79}$Se &     0.66 &   24( 7) &     6.96 & 0.16E+05(0.33E+04) &  0.97( 6) \\
$^{81}$Se &     0.60 &   12( 7) &     6.70 & 0.81E+04(0.35E+04) &  0.93(10) \\
$^{83}$Se &     0.48 &   17( 8) &     5.82 & 0.39E+04(0.73E+03) &  0.98(10) \\
$^{80}$Br &     0.36 &   75(18) &     7.89 & 0.17E+06(0.21E+05) &  0.98( 3) \\
$^{82}$Br &     0.90 &   43(11) &     7.59 & 0.98E+05(0.18E+05) &  0.87( 4) \\
$^{79}$Kr &     0.84 &   44( 9) &     8.36 & 0.13E+06(0.37E+05) &  0.94( 4) \\
$^{81}$Kr &     0.96 &   41(14) &     7.88 & 0.11E+06(0.12E+05) &  0.88( 4) \\
$^{83}$Kr &     0.84 &   11( 3) &     7.46 & 0.80E+05(0.51E+05) &  0.75( 6) \\
$^{86}$Rb &     1.08 &   32(11) &     8.65 & 0.77E+05(0.16E+05) &  0.97( 5) \\
$^{88}$Rb &     0.30 &   12( 7) &     6.08 & 0.57E+04(0.16E+04) &  0.94(10) \\
$^{85}$Sr &     1.14 &   16( 4) &     8.53 & 0.83E+05(0.32E+05) &  0.86( 5) \\
$^{87}$Sr &     1.50 &    9( 4) &     8.43 & 0.34E+05(0.11E+05) &  0.85( 6) \\
$^{88}$Sr &     2.52 &   16( 9) &    11.11 & 0.60E+05(0.67E+04) &  1.04( 7) \\
$^{89}$Sr &     2.04 &   20( 9) &     6.37 & 0.25E+04(0.49E+03) &  0.90( 9) \\
$^{90}$Y  &     0.96 &    9( 4) &     6.86 & 0.15E+05(0.28E+04) &  0.80( 5) \\
$^{91}$Zr &     2.28 &   37(15) &     7.20 & 0.52E+04(0.10E+04) &  1.00( 9) \\
$^{92}$Zr &     1.44 &    8( 6) &     8.64 & 0.24E+05(0.68E+04) &  0.90( 9) \\
$^{93}$Zr &     1.80 &   16( 4) &     6.73 & 0.13E+05(0.36E+04) &  0.74( 4) \\
$^{95}$Zr &     1.80 &   18( 6) &     6.46 & 0.13E+05(0.25E+04) &  0.71( 4) \\
$^{97}$Zr &     1.20 &    8( 6) &     5.58 & 0.11E+05(0.30E+04) &  0.61( 7) \\
$^{94}$Nb &     1.14 &   61(13) &     7.23 & 0.98E+05(0.14E+05) &  0.82( 3) \\
$^{93}$Mo &     2.76 &  103(21) &     8.07 & 0.20E+05(0.38E+04) &  1.01( 5) \\
$^{95}$Mo &     0.90 &   17( 9) &     7.37 & 0.30E+05(0.97E+04) &  0.87( 7) \\
$^{96}$Mo &     2.58 &   44(11) &     9.15 & 0.86E+05(0.12E+05) &  0.87( 3) \\
$^{97}$Mo &     0.90 &   20( 5) &     6.82 & 0.35E+05(0.61E+04) &  0.80( 3) \\
$^{98}$Mo &     2.46 &   51(12) &     8.64 & 0.99E+05(0.15E+05) &  0.82( 3) \\
$^{99}$Mo &     0.78 &   35(10) &     5.93 & 0.34E+05(0.84E+04) &  0.75( 4) \\
$^{101}$Mo &     0.42 &   44(11) &     5.40 & 0.46E+05(0.56E+04) &  0.72( 3) \\
$^{100}$Tc &     0.42 &   75(14) &     6.76 & 0.28E+06(0.55E+05) &  0.77( 3) \\
$^{100}$Ru &     2.34 &   46( 9) &     9.67 & 0.24E+06(0.53E+05) &  0.86( 3) \\
$^{102}$Ru &     2.16 &   24( 4) &     9.22 & 0.46E+06(0.61E+05) &  0.72( 2) \\
$^{103}$Ru &     0.36 &   34( 8) &     6.23 & 0.64E+05(0.19E+05) &  0.78( 4) \\
$^{104}$Ru &     1.38 &   11( 6) &     8.91 & 0.11E+07(0.56E+06) &  0.65( 4) \\
$^{104}$Rh &     0.48 &  121(22) &     7.00 & 0.56E+06(0.81E+05) &  0.77( 2) \\
$^{105}$Pd &     0.66 &   33( 6) &     7.09 & 0.16E+06(0.25E+05) &  0.76( 2) \\
$^{106}$Pd &     2.52 &   56(11) &     9.56 & 0.65E+06(0.97E+05) &  0.75( 2) \\
$^{107}$Pd &     0.54 &   24(11) &     6.54 & 0.14E+06(0.48E+05) &  0.69( 5) \\
$^{108}$Pd &     2.16 &   23( 6) &     9.22 & 0.72E+06(0.13E+06) &  0.68( 2) \\
$^{109}$Pd &     0.48 &   46(14) &     6.15 & 0.19E+06(0.77E+05) &  0.68( 4) \\
$^{111}$Pd &     0.36 &   28( 7) &     5.75 & 0.10E+06(0.17E+05) &  0.66( 3) \\
$^{108}$Ag &     0.60 &  115(22) &     7.27 & 0.70E+06(0.15E+06) &  0.77( 3) \\
$^{110}$Ag &     0.48 &   87(19) &     6.81 & 0.98E+06(0.20E+06) &  0.68( 2) \\
$^{107}$Cd &     0.66 &   14( 4) &     7.93 & 0.31E+06(0.85E+05) &  0.73( 3) \\
$^{109}$Cd &     0.96 &   26( 7) &     7.35 & 0.34E+06(0.91E+05) &  0.67( 3) \\
$^{111}$Cd &     0.84 &   38( 9) &     6.98 & 0.26E+06(0.42E+05) &  0.70( 2) \\
$^{112}$Cd &     2.88 &  190(57) &     9.40 & 0.86E+06(0.16E+06) &  0.77( 3) \\
$^{113}$Cd &     1.08 &   87(19) &     6.54 & 0.20E+06(0.44E+05) &  0.71( 3) \\
$^{114}$Cd &     1.86 &   19(10) &     9.04 & 0.83E+06(0.13E+06) &  0.67( 3) \\
$^{115}$Cd &     0.60 &   20( 6) &     6.14 & 0.17E+06(0.30E+05) &  0.61( 2) \\
$^{117}$Cd &     0.24 &   24(10) &     5.77 & 0.10E+06(0.25E+05) &  0.66( 4) \\
$^{114}$In &     1.08 &   68(16) &     7.28 & 0.10E+07(0.15E+06) &  0.65( 2) \\
$^{116}$In &     0.90 &   80(15) &     6.78 & 0.73E+06(0.52E+05) &  0.65( 1) \\
$^{113}$Sn &     1.14 &   19( 7) &     7.75 & 0.28E+06(0.95E+05) &  0.69( 4) \\
$^{115}$Sn &     0.60 &   12( 7) &     7.55 & 0.19E+06(0.68E+05) &  0.72( 5) \\
$^{116}$Sn &     2.22 &   16( 7) &     9.56 & 0.54E+06(0.27E+06) &  0.70( 5) \\
$^{117}$Sn &     1.56 &   44(17) &     6.95 & 0.11E+06(0.40E+05) &  0.69( 5) \\
$^{118}$Sn &     2.22 &   27(11) &     9.33 & 0.53E+06(0.13E+06) &  0.72( 3) \\
$^{119}$Sn &     1.08 &   18( 6) &     6.49 & 0.70E+05(0.19E+05) &  0.66( 3) \\
$^{120}$Sn &     1.74 &   12( 4) &     9.11 & 0.38E+06(0.83E+05) &  0.71( 3) \\
$^{121}$Sn &     1.02 &   24( 8) &     6.17 & 0.42E+05(0.13E+05) &  0.69( 4) \\
$^{123}$Sn &     1.02 &   28(13) &     5.95 & 0.26E+05(0.13E+05) &  0.72( 7) \\
$^{125}$Sn &     0.96 &   20( 9) &     5.73 & 0.17E+05(0.82E+04) &  0.71( 7) \\
$^{122}$Sb &     0.24 &   62(16) &     6.81 & 0.53E+06(0.72E+05) &  0.73( 2) \\
$^{124}$Sb &     0.48 &   61(16) &     6.47 & 0.34E+06(0.41E+05) &  0.69( 2) \\
$^{123}$Te &     0.60 &   25(10) &     6.93 & 0.32E+06(0.12E+06) &  0.67( 4) \\
$^{124}$Te &     2.76 &  112(22) &     9.43 & 0.14E+07(0.34E+06) &  0.71( 2) \\
$^{125}$Te &     0.54 &   24( 8) &     6.57 & 0.20E+06(0.50E+05) &  0.67( 3) \\
$^{126}$Te &     2.28 &   46(10) &     9.11 & 0.66E+06(0.83E+05) &  0.71( 2) \\
$^{127}$Te &     0.72 &   29(13) &     6.29 & 0.74E+05(0.14E+05) &  0.71( 4) \\
$^{129}$Te &     0.60 &   16( 5) &     6.09 & 0.22E+05(0.70E+04) &  0.75( 5) \\
$^{131}$Te &     0.96 &   25(10) &     5.93 & 0.79E+04(0.13E+04) &  0.86( 7) \\
$^{128}$I  &     0.30 &   75(14) &     6.83 & 0.52E+06(0.14E+06) &  0.74( 3) \\
$^{130}$I  &     0.24 &   47(14) &     6.46 & 0.32E+06(0.62E+05) &  0.71( 3) \\
$^{125}$Xe &     0.72 &   45( 9) &     7.60 & 0.76E+06(0.83E+05) &  0.71( 2) \\
$^{129}$Xe &     0.42 &   24( 8) &     6.91 & 0.18E+06(0.21E+05) &  0.73( 3) \\
$^{130}$Xe &     2.64 &   58(15) &     9.26 & 0.82E+06(0.11E+06) &  0.69( 2) \\
$^{131}$Xe &     0.84 &   23( 7) &     6.62 & 0.25E+06(0.82E+05) &  0.62( 3) \\
$^{132}$Xe &     2.16 &   29(11) &     8.94 & 0.35E+06(0.13E+06) &  0.72( 4) \\
$^{133}$Xe &     0.96 &   20( 9) &     6.45 & 0.63E+05(0.65E+04) &  0.68( 4) \\
$^{136}$Xe &     2.52 &   29(11) &     7.99 & 0.27E+05(0.60E+04) &  0.80( 5) \\
$^{134}$Cs &     0.42 &   62(10) &     6.89 & 0.44E+06(0.54E+05) &  0.73( 2) \\
$^{135}$Cs &     0.36 &    8( 6) &     8.83 & 0.68E+06(0.13E+06) &  0.75( 5) \\
$^{137}$Cs &     1.26 &    9( 3) &     8.27 & 0.15E+06(0.32E+05) &  0.72( 3) \\
$^{131}$Ba &     0.66 &   28( 9) &     7.49 & 0.76E+06(0.11E+06) &  0.67( 2) \\
$^{133}$Ba &     0.72 &   18( 6) &     7.19 & 0.37E+06(0.97E+05) &  0.65( 3) \\
$^{135}$Ba &     1.08 &   25( 7) &     6.97 & 0.13E+06(0.37E+05) &  0.69( 3) \\
$^{136}$Ba &     2.22 &   40(10) &     9.11 & 0.41E+06(0.11E+06) &  0.75( 3) \\
$^{137}$Ba &     0.48 &    4( 3) &     6.90 & 0.55E+05(0.13E+05) &  0.68( 5) \\
$^{138}$Ba &     1.80 &   11( 5) &     8.61 & 0.71E+05(0.24E+05) &  0.78( 5) \\
$^{139}$Ba &     2.04 &   62(16) &     4.72 & 0.31E+04(0.93E+03) &  0.69( 7) \\
$^{139}$La &     1.38 &   28( 9) &     8.78 & 0.42E+06(0.63E+05) &  0.77( 3) \\
$^{140}$La &     1.20 &   92(21) &     5.16 & 0.35E+05(0.28E+04) &  0.67( 3) \\
$^{137}$Ce &     1.20 &   29(11) &     7.48 & 0.52E+06(0.85E+05) &  0.64( 3) \\
$^{141}$Ce &     2.10 &   47(11) &     5.43 & 0.16E+05(0.54E+04) &  0.57( 4) \\
$^{142}$Pr &     0.90 &   34( 7) &     5.84 & 0.16E+06(0.39E+05) &  0.59( 2) \\
$^{143}$Nd &     2.04 &   83(25) &     6.12 & 0.75E+05(0.11E+05) &  0.60( 3) \\
$^{144}$Nd &     2.64 &   78(18) &     7.82 & 0.24E+06(0.32E+05) &  0.64( 2) \\
$^{145}$Nd &     1.26 &   41(11) &     5.76 & 0.91E+05(0.14E+05) &  0.58( 2) \\
$^{146}$Nd &     1.92 &   50(10) &     7.57 & 0.53E+06(0.61E+05) &  0.61( 2) \\
$^{147}$Nd &     0.84 &   37(12) &     5.29 & 0.15E+06(0.23E+05) &  0.54( 2) \\
$^{148}$Nd &     1.14 &   17( 7) &     7.33 & 0.21E+07(0.85E+06) &  0.53( 3) \\
$^{149}$Nd &     0.24 &   44(13) &     5.04 & 0.29E+06(0.46E+05) &  0.55( 2) \\
$^{151}$Nd &     0.60 &   50(14) &     5.34 & 0.29E+06(0.43E+05) &  0.55( 2) \\
$^{148}$Pm &     0.48 &   71(17) &     5.89 & 0.16E+07(0.44E+06) &  0.54( 2) \\
$^{148}$Sm &     1.62 &   22( 4) &     8.14 & 0.16E+07(0.18E+06) &  0.58( 1) \\
$^{150}$Sm &     2.04 &   79(18) &     7.99 & 0.37E+07(0.52E+06) &  0.55( 1) \\
$^{151}$Sm &     0.36 &   92(20) &     5.60 & 0.75E+06(0.13E+06) &  0.58( 2) \\
$^{152}$Sm &     1.14 &   31( 9) &     8.26 & 0.73E+07(0.12E+07) &  0.58( 2) \\
$^{153}$Sm &     0.30 &   68(14) &     5.87 & 0.88E+06(0.31E+06) &  0.59( 3) \\
$^{155}$Sm &     0.24 &   25(10) &     5.81 & 0.41E+06(0.57E+05) &  0.57( 3) \\
$^{152}$Eu &     0.24 &  307(36) &     6.31 & 0.12E+08(0.17E+07) &  0.57( 1) \\
$^{153}$Eu &     0.42 &   47(11) &     8.55 & 0.36E+08(0.29E+07) &  0.60( 1) \\
$^{154}$Eu &     0.36 &  296(35) &     6.44 & 0.88E+07(0.15E+07) &  0.59( 1) \\
$^{155}$Eu &     0.90 &   54(12) &     8.15 & 0.12E+08(0.28E+07) &  0.59( 2) \\
$^{156}$Eu &     0.24 &   78(18) &     6.34 & 0.24E+07(0.28E+06) &  0.59( 2) \\
$^{153}$Gd &     0.24 &   75(18) &     6.25 & 0.36E+07(0.60E+06) &  0.56( 2) \\
$^{155}$Gd &     0.36 &   78(18) &     6.44 & 0.36E+07(0.50E+06) &  0.57( 1) \\
$^{156}$Gd &     1.62 &   59( 8) &     8.54 & 0.10E+08(0.16E+07) &  0.57( 1) \\
$^{157}$Gd &     0.72 &  109(21) &     6.36 & 0.13E+07(0.19E+06) &  0.60( 2) \\
$^{158}$Gd &     1.26 &   38(10) &     7.94 & 0.32E+07(0.40E+06) &  0.59( 2) \\
$^{159}$Gd &     0.72 &   66(17) &     5.94 & 0.64E+06(0.70E+05) &  0.57( 2) \\
$^{161}$Gd &     0.36 &   25(10) &     5.64 & 0.27E+06(0.38E+05) &  0.57( 3) \\
$^{160}$Tb &     0.36 &   78(19) &     6.26 & 0.46E+07(0.20E+07) &  0.54( 2) \\
$^{159}$Dy &     0.48 &   56(15) &     6.83 & 0.20E+07(0.44E+06) &  0.61( 2) \\
$^{161}$Dy &     0.60 &   78(18) &     6.45 & 0.21E+07(0.25E+06) &  0.57( 1) \\
$^{162}$Dy &     1.98 &  128(19) &     8.20 & 0.60E+07(0.36E+07) &  0.58( 3) \\
$^{163}$Dy &     0.36 &   53(15) &     6.27 & 0.90E+06(0.15E+06) &  0.61( 2) \\
$^{164}$Dy &     0.96 &   20(10) &     7.66 & 0.23E+07(0.42E+06) &  0.58( 3) \\
$^{165}$Dy &     0.60 &   54(15) &     5.72 & 0.33E+06(0.59E+05) &  0.59( 2) \\
$^{166}$Ho &     0.48 &  112(22) &     6.24 & 0.38E+07(0.75E+06) &  0.55( 2) \\
$^{163}$Er &     0.36 &   49(14) &     6.90 & 0.76E+07(0.13E+07) &  0.55( 2) \\
$^{165}$Er &     0.48 &   71(17) &     6.65 & 0.26E+07(0.43E+06) &  0.59( 2) \\
$^{167}$Er &     0.84 &   87(19) &     6.44 & 0.14E+07(0.20E+06) &  0.58( 2) \\
$^{168}$Er &     1.32 &   49(15) &     7.77 & 0.34E+07(0.61E+06) &  0.58( 2) \\
$^{169}$Er &     0.60 &   30( 9) &     6.00 & 0.63E+06(0.92E+05) &  0.54( 2) \\
$^{171}$Er &     0.36 &   33(12) &     5.68 & 0.39E+06(0.55E+05) &  0.57( 2) \\
$^{170}$Tm &     0.72 &  154(25) &     6.59 & 0.56E+07(0.11E+07) &  0.56( 1) \\
$^{169}$Yb &     0.84 &   75(18) &     6.87 & 0.28E+07(0.32E+06) &  0.57( 1) \\
$^{171}$Yb &     0.96 &   83(19) &     6.62 & 0.27E+07(0.45E+06) &  0.54( 1) \\
$^{172}$Yb &     1.68 &   89(14) &     8.02 & 0.57E+07(0.74E+06) &  0.57( 1) \\
$^{173}$Yb &     0.54 &   35(10) &     6.37 & 0.99E+06(0.11E+06) &  0.57( 2) \\
$^{174}$Yb &     1.74 &   63(13) &     7.46 & 0.16E+07(0.22E+06) &  0.57( 1) \\
$^{175}$Yb &     1.08 &   62(16) &     5.82 & 0.38E+06(0.56E+05) &  0.55( 2) \\
$^{177}$Yb &     0.30 &   19( 7) &     5.57 & 0.33E+06(0.48E+05) &  0.54( 2) \\
$^{176}$Lu &     0.72 &  156(25) &     6.29 & 0.51E+07(0.82E+06) &  0.54( 1) \\
$^{177}$Lu &     0.72 &   41(13) &     7.07 & 0.80E+07(0.12E+07) &  0.52( 2) \\
$^{175}$Hf &     0.30 &   39(10) &     6.71 & 0.41E+07(0.86E+06) &  0.55( 2) \\
$^{177}$Hf &     0.54 &   43(11) &     6.38 & 0.20E+07(0.48E+06) &  0.54( 2) \\
$^{178}$Hf &     1.56 &   87(19) &     7.63 & 0.44E+07(0.64E+06) &  0.56( 1) \\
$^{179}$Hf &     0.54 &   36(10) &     6.10 & 0.12E+07(0.21E+06) &  0.53( 2) \\
$^{180}$Hf &     1.26 &   53(12) &     7.39 & 0.25E+07(0.31E+06) &  0.57( 1) \\
$^{181}$Hf &     0.24 &   20( 9) &     5.70 & 0.62E+06(0.26E+06) &  0.53( 3) \\
$^{181}$Ta &     0.24 &   17( 9) &     7.58 & 0.11E+08(0.96E+06) &  0.55( 2) \\
$^{182}$Ta &     0.48 &   85(19) &     6.06 & 0.25E+07(0.26E+06) &  0.54( 1) \\
$^{183}$Ta &     0.48 &   17( 8) &     6.93 & 0.33E+07(0.69E+06) &  0.53( 2) \\
$^{181}$W  &     0.48 &   49(14) &     6.68 & 0.37E+07(0.91E+06) &  0.55( 2) \\
$^{183}$W  &     0.42 &   28( 9) &     6.19 & 0.13E+07(0.34E+06) &  0.54( 2) \\
$^{184}$W  &     1.32 &   46(14) &     7.41 & 0.29E+07(0.37E+06) &  0.55( 2) \\
$^{185}$W  &     0.24 &   25(11) &     5.75 & 0.74E+06(0.10E+06) &  0.54( 2) \\
$^{187}$W  &     0.24 &   25(10) &     5.47 & 0.51E+06(0.62E+05) &  0.53( 2) \\
$^{186}$Re &     0.36 &   71(17) &     6.18 & 0.41E+07(0.51E+06) &  0.53( 1) \\
$^{188}$Re &     0.24 &   73(17) &     5.87 & 0.38E+07(0.54E+06) &  0.52( 1) \\
$^{187}$Os &     0.42 &   47(11) &     6.29 & 0.23E+07(0.50E+06) &  0.54( 2) \\
$^{188}$Os &     1.86 &   68(13) &     7.99 & 0.83E+07(0.89E+06) &  0.52( 1) \\
$^{189}$Os &     0.60 &   62(16) &     5.92 & 0.17E+07(0.18E+06) &  0.52( 1) \\
$^{190}$Os &     1.98 &   85(15) &     7.79 & 0.58E+07(0.64E+06) &  0.52( 1) \\
$^{191}$Os &     0.36 &   58(15) &     5.76 & 0.95E+06(0.12E+06) &  0.56( 2) \\
$^{193}$Os &     0.36 &   20( 9) &     5.58 & 0.58E+06(0.76E+05) &  0.51( 2) \\
$^{192}$Ir &     0.24 &  141(24) &     6.21 & 0.59E+07(0.77E+06) &  0.56( 1) \\
$^{193}$Ir &     0.48 &   41(14) &     7.76 & 0.19E+08(0.34E+07) &  0.56( 2) \\
$^{194}$Ir &     0.24 &   78(18) &     6.07 & 0.25E+07(0.35E+06) &  0.56( 2) \\
$^{195}$Pt &     0.48 &   54(15) &     6.11 & 0.29E+06(0.96E+05) &  0.66( 3) \\
$^{196}$Pt &     1.92 &   75(18) &     7.92 & 0.21E+07(0.41E+06) &  0.59( 2) \\
$^{197}$Pt &     0.48 &   37(12) &     5.85 & 0.18E+06(0.51E+05) &  0.63( 3) \\
$^{198}$Au &     0.42 &   65(13) &     6.51 & 0.13E+07(0.12E+06) &  0.62( 1) \\
$^{200}$Hg &     1.86 &   43( 8) &     8.03 & 0.47E+06(0.11E+06) &  0.66( 2) \\
$^{201}$Hg &     0.48 &   25(10) &     6.23 & 0.56E+05(0.70E+04) &  0.75( 4) \\
$^{202}$Hg &     1.74 &   41(11) &     7.76 & 0.18E+06(0.38E+05) &  0.72( 3) \\
$^{203}$Hg &     0.66 &   15( 6) &     5.93 & 0.29E+05(0.15E+05) &  0.70( 6) \\
$^{204}$Tl &     0.36 &   29(11) &     6.66 & 0.11E+06(0.19E+05) &  0.76( 4) \\
$^{206}$Tl &     1.02 &   15( 4) &     6.50 & 0.10E+05(0.40E+04) &  0.84( 6) \\
$^{205}$Pb &     1.68 &   44(13) &     6.73 & 0.29E+05(0.70E+04) &  0.78( 5) \\
$^{207}$Pb &     0.72 &    4( 3) &     6.74 & 0.33E+04(0.63E+03) &  0.91(10) \\
$^{208}$Pb &     3.42 &   26( 8) &     7.37 & 0.21E+04(0.68E+03) &  0.91( 9) \\
$^{209}$Pb &     1.08 &    5( 3) &     3.94 & 0.18E+03(0.78E+02) &  0.78(19) \\
$^{210}$Bi &     0.42 &   24( 8) &     4.61 & 0.23E+04(0.33E+03) &  0.92( 7) \\
$^{227}$Ra &     0.30 &   52(12) &     4.56 & 0.15E+07(0.35E+06) &  0.42( 1) \\
$^{230}$Th &     0.96 &   44(13) &     6.79 & 0.31E+08(0.13E+08) &  0.43( 2) \\
$^{231}$Th &     0.24 &   59(15) &     5.12 & 0.81E+07(0.14E+07) &  0.41( 1) \\
$^{233}$Th &     0.36 &   75(18) &     4.79 & 0.40E+07(0.13E+07) &  0.41( 2) \\
$^{235}$Pa &     0.24 &   20( 9) &     6.09 & 0.12E+09(0.21E+08) &  0.38( 1) \\
$^{233}$U  &     0.36 &   54(15) &     5.74 & 0.18E+08(0.33E+07) &  0.42( 1) \\
$^{234}$U  &     1.08 &   78(18) &     6.84 & 0.31E+08(0.69E+07) &  0.45( 1) \\
$^{235}$U  &     0.30 &   59(13) &     5.30 & 0.76E+07(0.37E+07) &  0.43( 2) \\
$^{236}$U  &     1.08 &   92(20) &     6.55 & 0.32E+08(0.83E+07) &  0.43( 1) \\
$^{237}$U  &     0.60 &   79(18) &     5.13 & 0.52E+07(0.61E+06) &  0.41( 1) \\
$^{238}$U  &     1.08 &   71(17) &     6.15 & 0.12E+08(0.29E+07) &  0.42( 1) \\
$^{239}$U  &     0.84 &   75(18) &     4.81 & 0.38E+07(0.52E+06) &  0.37( 1) \\
$^{238}$Np &     0.24 &   82(18) &     5.49 & 0.32E+08(0.29E+07) &  0.41( 1) \\
$^{239}$Pu &     0.42 &   49(11) &     5.65 & 0.94E+07(0.12E+07) &  0.43( 1) \\
$^{240}$Pu &     1.08 &   58(15) &     6.53 & 0.19E+08(0.20E+07) &  0.43( 1) \\
$^{241}$Pu &     0.30 &   46(11) &     5.24 & 0.60E+07(0.62E+06) &  0.42( 1) \\
$^{242}$Pu &     1.08 &   44(13) &     6.31 & 0.17E+08(0.25E+07) &  0.41( 1) \\
$^{243}$Pu &     0.30 &   26( 8) &     5.03 & 0.46E+07(0.72E+06) &  0.39( 1) \\
$^{242}$Am &     0.24 &   65(16) &     5.54 & 0.33E+08(0.72E+07) &  0.40( 1) \\
$^{243}$Am &     0.30 &   38(10) &     6.36 & 0.77E+08(0.21E+08) &  0.42( 1) \\
$^{244}$Am &     0.36 &  106(20) &     5.37 & 0.27E+08(0.52E+07) &  0.40( 1) \\
$^{243}$Cm &     0.84 &   32(11) &     5.70 & 0.67E+07(0.16E+07) &  0.40( 1) \\
$^{244}$Cm &     1.08 &   32(11) &     6.80 & 0.20E+08(0.31E+07) &  0.43( 1) \\
$^{245}$Cm &     0.48 &   58(15) &     5.52 & 0.71E+07(0.92E+06) &  0.43( 1) \\
$^{246}$Cm &     1.02 &   36(10) &     6.46 & 0.23E+08(0.10E+08) &  0.41( 2) \\
$^{247}$Cm &     0.36 &   61(16) &     5.16 & 0.26E+07(0.48E+06) &  0.45( 1) \\
$^{248}$Cm &     1.08 &   32(11) &     6.21 & 0.82E+07(0.19E+07) &  0.41( 1) \\
$^{249}$Cm &     0.24 &   32(14) &     4.71 & 0.25E+07(0.44E+06) &  0.40( 2) \\
$^{250}$Bk &     0.24 &   97(20) &     4.97 & 0.13E+08(0.17E+07) &  0.40( 1) \\
$^{250}$Cf &     1.08 &   38( 8) &     6.62 & 0.17E+08(0.27E+07) &  0.43( 1) \\
$^{253}$Cf &     0.24 &   12( 7) &     4.81 & 0.31E+07(0.55E+06) &  0.37( 2) \\
\end{tabular}
\end{table}

\newpage


\end{document}